\documentclass[aps,prb,twocolumn,showpacs]{revtex4}
\usepackage{epsfig}
\usepackage{color}
\newcommand{\boldsymbol}{}
\newcommand{\be}{\begin{equation}}
\newcommand{\ee}{\end{equation}}
\newcommand{\bea}{\begin{eqnarray}}
\newcommand{\eea}{\end{eqnarray}}

\newcommand{\remark}{}%{{\bf\large ** }}
\newcommand{\gergo}{}%{\bf}
\newcommand{\akos}[1]{}
\newcommand{\bff}{}%{}
\newcommand{\gf}{}%{}
\newcommand{\comma}{,}
\begin{document}
\title{Dynamical correlations and quantum phase transition in the %{\gf $Q=3$-state} 
quantum Potts model}
\author{\'Akos  Rapp$^{1,2}$ and Gergely Zar\'and$^{1,2}$}
\affiliation{
$^{1}$ Institut f\"ur Theoretische Festk\"orperphysik, Universit\"at Karlsruhe\comma  76128 Karlsruhe\comma Germany  \\
$^{2}$Theoretical Physics Department, Institute of Physics, Technical University Budapest,Budapest, H-1521, Hungary  }
\date{\today}
\begin{abstract}
We present a detailed study of the finite temperature dynamical properties 
of the quantum Potts model in one dimension.
Quasiparticle excitations in this model have 
internal quantum numbers, and their
scattering matrix {\gf deep} in the gapped phases is shown to take a simple {\gf exchange}  form in the
perturbative regimes.  The finite temperature
correlation functions in the quantum critical regime are
determined using conformal invariance, while {\gf far from the quantum critical point}
we compute the decay functions analytically within a semiclassical approach of 
Sachdev and Damle [K. Damle and S. Sachdev,  Phys. Rev. B \textbf{57}, 8307 (1998)]. 
As a consequence, decay functions exhibit a
{\em diffusive character}. 
{\gf We also provide robust arguments that our semiclassical analysis carries over to very low temperatures 
even in the vicinity of the quantum phase transition.}
Our results are also relevant for
quantum rotor models, antiferromagnetic chains,  and some spin ladder systems.
\end{abstract}
\maketitle

\section{Introduction}

Quantum critical systems have been in the focus of intense studies in the last
two decades.\cite{Sachdev_book,osszefogl} In these systems a second order phase transition
driven by quantum  fluctuations takes place at $T=0$ temperature, which,
nevertheless, determines the physical properties of the
system over a finite parameter range, and leads to anomalous dynamical scaling.
Much of our interest in these quantum critical systems has been motivated by the
anomalous scaling behavior that can be observed in various heavy fermion
compounds,\cite{AndrakaTsvelik,Schroeder,Aronson,HF_quantum_criticality}
but the anomalous behavior of the normal states of cuprates has also been
interpreted in terms of an underlying quantum phase transition, possibly
hidden by the d-wave superconducting state, similar to many heavy fermion 
compounds.\cite{varma,sachdev_QC,HF_supercond_quantumcrit}
Much of the studies of quantum critical phase transitions focused
on the simplest case of {\em magnetic} phase
transitions,\cite{Sachdev_book,Millis,Georges,sachdev_young} where typically  quantum
fluctuations compete with a tendency of magnetic ordering. While
in many cases a quantum-classical mapping enables one to map out
the phase diagram and determine critical exponents based upon our
knowledge and experience with classical criticality, computing
real time response functions represents a major challenge for
theorists. In this regard 1+1-dimensional systems are of crucial
importance since powerful methods can be used there to analyze
their dynamical properties both in the gapped
phases,\cite{sachdev_young,tsvelik_controzzi,Zotos} and in the quantum
critical regime \cite{Sachdev_book}. These one-dimensional models, 
besides being relevant to some experimental  systems,\cite{1D_systems}  serve also as
test grounds for higher-dimensional systems.\cite{bitko}

Maybe the simplest and most thoroughly studied 1+1-dimensional model is the {\em transverse field
Ising model}, where a one-dimensional Ising chain with ferromagnetic coupling $J$
is coupled to a magnetic field $h$ in the $x$ direction. In this model a quantum phase transition
takes place from a ferromagnetically ordered to a paramagnetic state at a critical value of the
magnetic field, $h_c/J=1$, and the critical theory is simply that of the two-dimensional
classical
Ising model.\cite{Sachdev_book,Ising_model,DiFrancesco} In this model, dynamical properties 
are well-understood both in the critical regime and in the gapped 
phases.\cite{Sachdev_book,sachdev_young}

In case of the transverse field Ising model, our deep
understanding of the quantum critical regime is based on the
observation that  a 1+1-dimensional critical system is
conformally invariant, and one can  therefore describe it 
by means of some conformal field theory.\cite{DiFrancesco} More interestingly, however,
we can turn this observation backwards by saying that any
conformal field theory should correspond to some model that
displays a quantum phase transition.

In this way, we can, in principle classify, construct, and study new
one-dimensional quantum mechanical systems which belong to {\em
different universality classes},  and thus exhibit new and interesting
critical behavior.

Relying on this observation, in the present paper we shall study the   simplest possible
1+1-dimensional generalization of the transverse  field Ising model which belongs to a
different universality class,  the $Q$-state quantum Potts model.  In this one-dimensional
model each spin has  $Q$ different components corresponding to the basis
states $|\mu\rangle$ with $\mu=1,\dots,Q$, and the Hamiltonian takes on the following
simple form

\begin{equation}\label{eq:defHamiltonipro}
{H}= - j\sum_{i} \sum_{\mu} {P}^{\mu}_i
{P}^{\mu}_{i+1} -jg\sum_{i} {P}_i,
\label{eq:H_Potts}
\end{equation}

where the operator $\boldsymbol{P}^{\mu}$ projects on state
$|\mu\rangle$ within the $Q$-dimensional local Hilbert space,
while $\boldsymbol{P}$ projects to the 'corner' state $\sum_\mu
|\mu\rangle/\sqrt{Q}$. The ferromagnetic coupling $j$ thus tries
to polarize all Potts spins at each site to one of the $Q$
possible orthogonal directions, while the magnetic field $h = j g$ tries to
project each spin to the $(1,1,\dots)$ direction and thus to
destroy long range order generated by the ferromagnetic coupling.
In the particular case of $Q=2$ Eq.~(\ref{eq:H_Potts}) reduces to
the Ising Hamiltonian, with $\sum_{\mu=1,2} {P}^{\mu}_i
{P}^{\mu}_{i+1} \equiv (\sigma^z_i \sigma^z_{i+1} + 1)/2$ and $P_i \equiv
(\sigma^x_i + 1) /2$. For later convenience, we shift the ground state energy 
and replace the projectors in Eq.~(\ref{eq:H_Potts}) by the  traceless operators
$\tilde{P}^{\mu}=P^{\mu}-\frac{1}{Q}$ and
$\tilde{P}=P-\frac{1}{Q}$.

The quantum Potts model defined by Eq.~(\ref{eq:H_Potts}) and the 
corresponding two-dimensional classical  Potts model have been studied 
extensively before.\cite{DiFrancesco,Potts,Potts_review,Potts_CFT,MittagStephen,Solyom} 
However, while we know  a lot about the quantum Potts model's thermodynamic 
behavior, much less is known about its {\em real time} 
dynamics, especially at finite temperatures.  The purpose of the present 
paper is to study in detail  {\em dynamical} correlations
within the quantum Potts model  at {\em finite temperatures}, both in the gapped and
in the quantum critical regimes. As we shall see, many of our results also 
apply to other one-dimensional gapped systems and are therefore 
of direct physical relevance. 

The one-dimensional quantum Potts model, Eq.~(\ref{eq:H_Potts}),
is known to exhibit a phase transition as a function
of the coupling $g$ \cite{MittagStephen,Solyom}: Below a critical value, $g<g_c$ the ground state
of Eq.~(\ref{eq:H_Potts}) is a $Q$-fold degenerate
{\em ferromagnet}   corresponding to the $Q$ different ferromagnetic
 alignments of the Potts  spins. The elementary excitations
are domain walls that move along the chain with a dispersion
\begin{equation}\label{eq:ferrogerj_energia}
\epsilon_{k}^{\mu,\mu'} = \epsilon(k)\;,
\end{equation}
where the two indices $\mu$ and $\mu'$ denote the orientations of the ferromagnetic order
parameters on the two sides of the excitation, and $k$ is its lattice momentum.
These quasiparticles are gapped and their energy can be approximated
for small  $g$-s as  $\epsilon(k)\approx j \left( 1 - g \;\frac{2}{Q}
\cos{k} \right)$.

For $g>g_c$, on the other hand, the ground state is a
non-degenerate {\em paramagnet}, which corresponds to orienting
all spins along the $(1,\dots,1)$ direction. Elementary
excitations of this state consist of  $\lambda=1,\dots,Q-1$
possible local  'spin flips', which propagate  along the chain
with a dispersion
\begin{equation}
\epsilon_{k}^{\lambda} = \tilde \epsilon(k) \;.
\label{eq:paragerj_energia}
\end{equation}
These excitations are also gapped and
for very large values of $g$ one finds $\tilde \epsilon(k)\approx
jg(1- \frac{2}{Q} \frac{1}{g}  \cos{k})$
from perturbation theory.

Clearly, since the structure of the ground state and its
elementary excitations  is \remark entirely different for $g\ll1$ and $g\gg1$,
a phase transition must occur at some critical value  $g=g_c$.
This phase transition turns out to be of second order for $Q\le 4$,
while it is  of first order for $Q>4$ \cite{Potts_review,Clock_model}. This
implies that the quasiparticle gap remains finite for $Q>4$, while
it approaches zero at $g=g_c$ for $Q\le 4$. The case $Q=4$ is special, 
and shall not be discussed here: Although the phase transition is continuous 
for $Q=4$ too, there a marginal operator dominates the critical point.\cite{marginalop} 
The case $Q=3$, on the other hand,  is of special interest: For $Q=3$ one finds a standard 
second order  quantum phase transition, which belongs to a universality class different from
the $Q=2$ Ising case \cite{DiFrancesco}. Furthermore, for $Q\ge 3$
the quasiparticles have  internal quantum numbers  making
their dynamics very interesting, and in many ways similar to that
of quantum rotor models and $S=1$ spin antiferromagnetic chains \cite{Sachdev_book,Damle}. 

\begin{figure}[h]
\centering
\includegraphics[width=8cm]{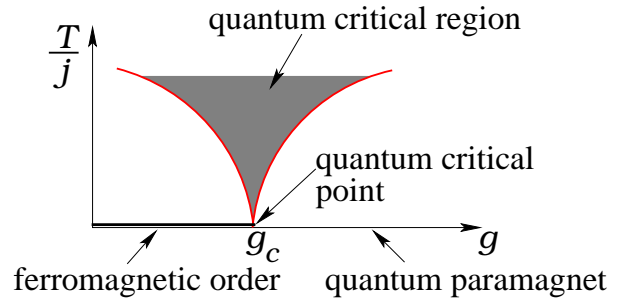}
\caption{\label{abra:fazisdiagram} (Color online) Sketch of the phase diagram of the one-dimensional $Q=3$ state quantum
Potts model. The quantum fluctuations become dominant at a temperature scale $T \sim \Delta$.}
\end{figure}

The phase diagram of the $Q=3$ quantum Potts model is sketched in
Fig.~\ref{abra:fazisdiagram}.

As we indicated in the figure, the quasiparticle gap $\Delta$ approaches zero as a power
law at the quantum critical point $g_c$,
\be
\Delta_{Q=3}(g) \approx C_\pm |g-g_c|^{5/6}\;,
\label{eq:gap}
\ee
where $C_\pm$ denote prefactors corresponding to the $g>g_c$ and $g<g_c$
regimes. The exponent in Eq.~(\ref{eq:gap})  simply follows from the quantum
classical mapping between the two-dimensional classical Potts model and
the one-dimensional quantum Potts model,\cite{Solyom,Kogut_review}  and is simply the 
critical exponent $\nu=5/6$ of the correlation length $\xi \sim 1/\Delta$ in 
the two-dimensional classical
problem.\cite{Sachdev_book,Nijs,DiFrancesco,Cardy} In the  quantum critical regime, $\Delta< T<j$,
physical properties of the model are governed by the
zero temperature critical point $g= g_c$.

The basic difference between the above two quantum phases can be
captured by the change in the dynamical structure factor
$S^{\mu\mu'}(\omega,q)$, defined as the Fourier transform of the
spin-spin correlation function, \remark
\be 
S_{\mu,\mu'} (t,x_i) \equiv
\left\langle \tilde P^{\mu}_i(t) \; \tilde P^{\mu'}_0(0)
\right\rangle\;. 
\ee 
The structure factor  $S^{\mu\mu'}(\omega,q)$ is directly  
measured by neutron scattering in magnetic systems, and it 
is related to the dynamical susceptibility, 

\be 
\chi_{\mu,\mu'} (t,x_i) \equiv (-i)
\left\langle [\tilde P^{\mu}_i(t)\;, \tilde P^{\mu'}_0(0)] 
\right\rangle\;\theta(t)\;,
\ee 

through the relation \remark %AKOS: check this

\be
S_{\mu\mu'}(\omega,q) = -2  (n(\omega) + 1)\; {\rm Im}\{\chi_{\mu,\mu'}(\omega,q)\}\;,
\label{eq:FDT}
\ee
with $n(\omega)$ the Bose function.

In the ferromagnetic phase at  $T=0$ temperature the  structure
factor has a Dirac delta component at $\omega = q = 0$\remark
\begin{eqnarray}
&& S^{T=0}_{\mu,\mu'}(\omega,q)  
=
(2\pi)^2 
\delta(\omega) \delta(q) \;m^2 \; \Bigl(\delta_{\mu,\tilde \mu} -
\frac{1}{Q} \Bigr)
\nonumber \\
&&
\phantom{nnnnnn}
\Bigl(\delta_{\mu',\tilde \mu} -
\frac{1}{Q} \Bigr) + \dots,
\label{eq:struktfakt_ferro_alapall}
\end{eqnarray}

where the order parameter $m$ is related to the expectation value of ${\tilde P}^{\mu}$ 
as $\langle{\tilde P}^{\mu}\rangle = m (\delta_{\mu,\tilde \mu} - 1/Q )$,
the order parameter being aligned along direction $\tilde \mu$.

\begin{figure}[h]
\centering
\includegraphics[width=6.5cm]{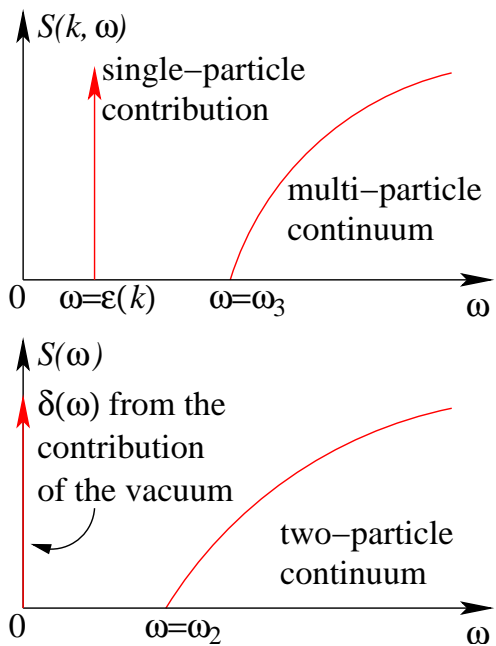}
\caption{\label{fig:S(omega)} (Color online) Schematics of the $T=0$ temperature
structure factors on the paramagnetic (top) and ferromagnetic (bottom)
sides of the phase transition. On the bottom we show the integrated 
structure factor, $S(\omega) \sim \int dq \; S(\omega,q)$.
Indices are dropped for simplicity.
}
\end{figure}

The $\omega=q=0$  delta function is a characteristic of long range order and is
absent for  $g>g_c$, where the structure factor
has a delta peak at the quasiparticle energy at $T=0$ temperature
\be
S^{T=0}_{\mu,\mu'}(\omega,q)
= A(g)\; \Bigl( \delta_{\mu \mu'} -1/Q \Bigr)\;
\delta(\omega - \tilde \epsilon(q))+\dots\;.
\label{eq:structure_factor_para}
\ee
This behavior is schematically shown in Fig.~\ref{fig:S(omega)}. For
$Q<4$ both the order parameter $m(g)$ and the quasiparticle residue
$A(g)$ scale to  zero as we approach the quantum critical point
$g_c$, where the quasiparticle gap $\Delta$ vanishes and the quasiparticle 
description breaks down.

The $T=0$ temperature dynamics discussed so far gets drastically modified in the
gapped phases at finite temperatures where  collisions between
quasiparticles must also be taken into account. Fortunately, we
can get an  accurate description of the low temperature
dynamics for $T\ll \Delta$ using the {\em semiclassical formalism}
of Refs.~[\onlinecite{tagging,sachdev_young,Damle}].

In this limit, we can describe
the spin system as a dilute gas of weakly interacting
quasiparticles with momenta $k\approx  0$.  Performing a
perturbative calculation we find that, {\bf in the perturbative regime}, for $Q\ne 4$ 
the scattering matrix ${\cal S}$ of
these  quasiparticles takes on a  ${\rm SU}(Q-1)$-symmetrical 
form in agreement with the general arguments  of
Ref.~[\onlinecite{Damle}]. In the paramagnetic phase we find, {\em
e.g.}, 
\be
{\cal S}^{\lambda',\tilde \lambda'}_{\lambda,\tilde \lambda}
\approx (-1) \; \delta^{\lambda'}_{\tilde \lambda} \delta^{\tilde
\lambda'}_{\lambda} \;, 
\label{eq:univ_s_matrix} 
\ee
thus quasiparticles exchange their internal quantum numbers under the
collision while the many-body wave function picks up a phase $(-1)$.

{\gf Far on the ferromagnetic side, $g\ll 1$}, 
we also find that the scattering matrix 
assumes the exchange form above, excepting that now \akos{quasiparticles}
can be viewed as {\em kinks} (domain walls) between different vacuum states. 
In course of the scattering process these kinks  exchange their quantum 
numbers in such a way that the the  domain in the ``middle'' of the 
collision keeps its color (see  Section~\ref{ferroside}).

{\gf Rather surprisingly, for $Q=3$ the results found in the perturbative regime 
 do not agree with the $\cal S$-matrix 
that one obtains by removing the cut-off and assuming 
that the system flows to a massive integrable fixed point \cite{Swieca}:
\be
 ({\cal S}_{\rm diag})^{\lambda',\tilde \lambda'}_{\lambda,\tilde \lambda}
\sim  (-1)^{(\lambda + \tilde \lambda)/2}\; \delta^{\lambda'}_{ \lambda} \delta^{\tilde
\lambda'}_{\tilde \lambda} \;. 
\label{eq:integr_s_matrix} 
\ee

One, rather unlikely explanation  is,  that Eq.~(\ref{eq:univ_s_matrix})
only describes the scattering of 
$q\to 0$ momentum quasiparticles far away from the quantum-critical point, and at 
for some finite values of 
$g\ne g_c$ 'phase transitions' occur where the asymptotic form of the quasiparticles 
jumps from one 'universal' form to the other.

 However, as we discuss in Appendix~\ref{appendix:RG}, rather robust arguments 
constructed along the lines of  Ref.~[\onlinecite{Damle}] suggest that  
Eq.~(\ref{eq:univ_s_matrix})  is indeed the $\cal S$-matrix of the $q\to0 $ 
momentum quasiparticles on a {\em lattice}: The low energy effective field theory 
contains three independent coupling constants, and unless two of these are the same, 
the $\cal S$-matrix takes the form (\ref{eq:univ_s_matrix}) rather than 
(\ref{eq:integr_s_matrix}). 

It is, nevertheless, quite concievable 
that the previously mentioned two couplings become equal 
once one {\em removes} the cut-off $\Lambda\sim j$. 

This in turn means  that for  $\Delta \ll j$
an intermediate energy regime may exist, where, rather than Eq.~(\ref{eq:univ_s_matrix}), 
 the diagonal $\cal S$-matrix (\ref{eq:integr_s_matrix}) of Ref.~\onlinecite{Swieca} describes 
the scattering processes. 

In fact, numerical calculations 
of the mass spectrum for $Q>3$ seem to support that  the above K\"oberle-Swieca $\cal S$-matrix 
correctly describes quasiparticles at large momenta, $q\sim \Delta/c$, where bound solitons are 
formed.\cite{Alcaraz}  

Therefore, while we believe that 
the correlation functions we derive based upon  Eq.~(\ref{eq:univ_s_matrix}) always  correctly 
describe the regime $T\to 0$ ($q\to 0 $), the range of validity of the results 
in the regime $|g-g_c|\ll 1$ needs further numerical investigation.}

{\gf
We remark that there are a number of cases where in the 'universal scaling limit'
important corrections are missed that actually provide the leading 
contributions to some quantities, so that} 
 the Potts model would not be unique at all in this regard: 

A famous example  
is provided by the so-called free fermion point of the  sine-Gordon model, 
where in the continuum limit one also finds a diagonal $\cal S$-matrix. 
There too, however, the asymptotic $\cal S$-matrix assumes the exchange form
immediately once one introduces a cut-off or goes slightly away from this special 
point.\cite{Sine_Gordon?} {\gf 
The same happens in the two-channel Kondo model
in the scaling limit. There the channel susceptibility and also that
associated with the composite superconducting order parameter simply {\em vanishes}
in the scaling limit, because the 
coefficients of these terms are inversely proportional to the cut-off 
\cite{CoxZawa}. There are also abundant examples in the literature where 
the physical properties of a model change dramatically 
once the integrability condition is violated (random matrix theory, 
integrable models with just a single impurity, {\em etc.}).
}

The above  simple form of the \akos{asymptotic quasiparticle} ${\cal S}$-matrix,
Eq.~(\ref{eq:univ_s_matrix}),
 then enables us to
compute {\em analytically} the correlation function within the
semiclassical approximation \akos{as $T \to 0$}. This turns out to be a universal
function of the typical separation $\xi_c$ between the
quasiparticles and their scattering time $\tau$, given explicitely by Eqs.~(\ref{eq:xi_c_def})
and (\ref{eq:tau}), respectively.

In the ferromagnetic phase we can express the spin-spin correlation
function for $T\to 0$ as
\be
\label{eq:korrfv_vegsoalak_ferro} 
S_{\mu,\mu'}(x,t) =
{m^2}\;\frac{1}{Q}\;
\big(\delta_{\mu,\mu'}-\frac 1Q
\big) \;R(\bar{x},\bar{t}) \;,
\ee

where the relaxation function $R$ is given by the following expression:

\bea
&&R(\bar x ,\bar t)=\int_{-\pi}^{\pi} \frac{d \phi}{2\pi} e^{-
| \bar{t} | \; (1-\cos{\phi}) \; \left( \frac{1}{\sqrt{\pi}}
e^{-u^2} + u \; {\rm erf}(u) \right) }
\phantom{nnnn}
\label{eq:relax_fv_integralalak}
\\
&&\phantom{nnn}
\cos{ \left(
\sin({\phi}) \; \bar{x} \right) } \; \frac{(Q-1)^2-1}{
(Q-1)^2+1+2(Q-1)\cos{\phi}}\;,
\nonumber
\eea

and $\bar x = x/\xi_c$, $\bar t = t/\tau$, and  $u = \bar x / \bar t$,
denote dimensionless separations, times and velocities.

This formula is quite remarkable: On one hand, in the $Q\to 2$ limit it reproduces the
exact results of Refs.~[\onlinecite{Sachdev_book,sachdev_young}]. However, while
for $Q=2$ the function $R$ decays exponentially, for $Q>2$ it 
has a {\em diffusive structure} for 
$1 \ll \bar t$:
\be
R(\bar x,\bar t) \sim \frac 1{\sqrt{\bar{t}}}
\;  {\rm exp} \left( -\frac{\sqrt{\pi}\; \bar x^2}{2 \bar{t} } \right) \;.
\ee
This diffusive structure is related to
the approximate ${\rm SU}(Q-1)$ invariance of the \akos{quasiparticle} scattering matrix.
As a consequence, the $T=0$ temperature Dirac delta peak in the structure 
factor is replaced by\remark
%This should be checked:

\bea
S_{\mu\mu'} (\omega,q) \sim m^2 \left(\delta_{\mu,\mu'}-\frac 1Q \right) \; 
%{ \xi_c \tau} 
\frac{\tau \xi_c^3 q^2 }{\xi_c^4 q^4 +\omega^2 \tau^2 \;{4 {\pi}} }
%\nonumber 
%\\ 
%\times %\left(
%\frac{\xi_c^2 q^2 }{\xi_c^4 q^4 + 4 \pi \omega^2 \tau^2 }
%+\frac{1}{\xi_c^2 q^2 -i\omega \tau \;{2  \sqrt{\pi}}} \right) 
\;.
\eea

From this, with the help of (\ref{eq:FDT}), we find that the 
susceptibility in the (semiclassical) limit, $\omega \ll T$ 
has a diffusion pole,

\be
\chi_{\mu\mu'} (\omega,q) \sim m^2 \left(\delta_{\mu,\mu'}-\frac 1Q \right) \frac{\xi_c}{T} \; \frac{\xi_c^2q^2}{\xi_c^2q^2-i\omega\tau2 \sqrt{\pi}} \;.
\ee

The prefactor in this expression, $\sim m^2 \xi_c /T$ is just the static susceptibility, 
that can be interpreted as the Curie susceptibility of independent domains of size $\xi_c$ 
having  magnetization $m$.

In fact, this diffusive structure is rather natural, and we shall discuss
a simple explanation for it later. What is not natural, is the
exponential decay found in the transverse field Ising model
($Q=2$).\cite{sachdev_young} As we discuss later, 
 this exponential decay is a consequence of long range correlations in the domain wall orientations
along the chain for  $Q=2$, and is a peculiarity of the Ising model.

For the low temperature correlation function on the paramagnetic side of the
phase transition we find the following expression \remark
\begin{equation}
S_{\mu,\mu'}(x,t) =
A\; \left(\delta_{\mu,\mu'} - \frac{1}{Q} \right) \; 
K(x,t) \; R(\bar x ,\bar t) \;,
\label{eq:korrfv_vegso_alak_para}
\end{equation}
where the propagator $K(x,t)$ is approximately
the Feynman propagator of a quantum mechanical particle of mass $\Delta$,
and is given by
\begin{equation}
K(x,t) \approx  \, e^{-i\Delta \; t} \,
\sqrt{\frac{\Delta}{2\pi i \;t}} \; {\rm exp}\left( i \frac{\Delta \;
x^2}{2 c^2 t} \right)
\label{eq:Feynmanprop_para}
\end{equation}
for $x$ and $t$ within the light cone.
Remarkably, the relaxation function $R(\bar x ,\bar t)$ is the {\em same }
as the one found  in the ferromagnetic phase. This
is very likely the  consequence of the self-duality of the $Q$-state
Potts model.\cite{Potts,MittagStephen}

{\gf We have to emphasize that Eqs.~(\ref{eq:relax_fv_integralalak}) and (\ref{eq:korrfv_vegso_alak_para})
rely on the structure~(\ref{eq:univ_s_matrix})
of the ${\cal S}$-matrix. Therefore, while they are certainly valid in the 
regime far from the critical point, $|g-g_c|\sim 1$, it is not clear below what temperature these formulas describe the correlation 
functions for $|g-g_c|\ll 1$. This issue needs some further numerical 
investigation which is beyond the scope of the  persent paper. 
}

It is important to emphasize 
that the results Eq.~(\ref{eq:korrfv_vegso_alak_para})
obtained above are also relevant for $S=1$ antiferromagnetic spin chains,\cite{Damle,Affleck} 
 and some two-leg ladder systems 
that can be mapped to a one-dimensional $O(3)$  quantum rotor model, 
also equivalent to the $O(3)$ sigma model in the long time and long wavelength 
limit.\cite{Sachdev_book,Haldane}  The one-dimensional quantum rotor model 
consists of a chain of ferromagnetically coupled quantum rotors, and is 
defined as 

\be 
H_{\rm rotor} =  g \sum_i {J\over 2} {\vec L}_i^2 - J \sum_i {\vec n}_i {\vec n}_{i+1}\;.
\ee

 Here the vectors ${\vec n}_i$ denote $N$-dimensional unit
vectors and the ${\vec L}_i$'s denote the corresponding angular
momentum operators. This  model maps to the  two-dimensional classical $O(N)$
model. Correspondingly, it does not display a quantum phase
transition but has only a paramagnetic phase, where the coupling
$g$ always generates a finite gap.\cite{Sachdev_book} On the other hand, this
paramagnetic phase is similar to that of the $Q$-state Potts model
in that the quasiparticles have internal quantum numbers. Let
us focus here  on the probably most relevant $N=3$ case, the $O(3)$
rotor model, where the gapped quasiparticles are spin $S=1$  objects.
Since the scattering matrix takes on the same universal form as in
our case\cite{Damle}, all our calculations of the finite temperature
properties carry over to this case as well and give 
\be 
\langle {\vec n}(x,t)\cdot {\vec n}(0)\rangle = A\;  K(x,t) \; R_{Q=4}(\bar x ,\bar t) \;, 
\ee 
where $R_{Q=4}$ is just the relaxation function Eq.~(\ref{eq:relax_fv_integralalak})
with $Q=4$.

Thus our results are also relevant to  spin $S=1$
antiferromagnetic chains and also some of the
experimentally studied spin ladder systems,\cite{1D_systems}
for which  the one-dimensional $O(3)$ quantum rotor model provides
a satisfactory description of low energy (long wavelength)
fluctuations.\cite{Haldane,halperin}

The semiclassical formalism discussed above  gives a
consistent description of the dynamical  fluctuations at very low temperatures in the
gapped phases. Conceptually, however,  the most interesting regime is
the quantum critical regime, $\Delta < T < j$. In
this regime dynamical correlations are governed by fluctuations 
related to the  critical point, $g= g_c$, and can be accessed by making use of the
conformal invariance of the critical theory. Conformal theory
implies that the  finite temperature dynamical susceptibility is given 
approximately by \remark
\bea
\chi^{T\gg \Delta}_{\mu\mu'}(\omega,q) &\sim & \big(\delta_{\mu,\mu'}-\frac 1Q \big)
{1\over T^{26/15}} 
\label{eq:dynsus_qcrit}
\\
&\times &\; \frac{\Gamma(\frac{1}{15}-i\frac{\omega+cq}{4\pi T}) \; \Gamma(\frac{1}{15}-i\frac{\omega-cq}{4\pi T})}{\Gamma(\frac{14}{15}-i\frac{\omega+cq}{4\pi T}) \; \Gamma(\frac{14}{15}-i\frac{\omega-cq}{4\pi T})}
\nonumber 
\eea
for $\omega,cq, T >\Delta$, and correspondingly, the susceptibility 
exhibits $\omega/T$ scaling in this quantum critical regime. A similar scaling form 
shall be obtained for the local susceptibility.

{\bff Although we think they are irrelevant {\gf for $T\to 0$}, for completeness, let us also 
give here the relaxation functions obtained under {\gf the  
assumption of integrability} in the gapped phases, 
{\em i.e.} using the diagonal $\cal S$-matrix, Eq.~(\ref{eq:integr_s_matrix}). In this case we obtain 
the expression
\bea
R_{\rm diag}^{\rm para}(\bar x,\bar t) &=&  e^{-\bar t G(\bar x/\bar t)}\;,\\
G(u) &=&  \frac{1}{\sqrt{\pi}} e^{-u^2} + u \;{\rm erf}( u)\;
\label{eq:R_int^para}
\eea
for the decay function in the paramagnetic case, while in the ferromagnetic case
we find 

\be
R_{\rm diag}^{\rm ferro}(\bar x,\bar t) =  e^{-\frac 32 \;\bar t G(\bar x/\bar t)}\;.
\label{eq:R_int^ferro}
\ee

In both cases the correlation function decays exponentially. 
The derivation of these expressions, {\gf which may be relevant in an intermediate temperature range, 
$\Delta \gg T > T^*$ for $|g-g_c|\ll 1$,}  
is given in Appendix~\ref{app:integr}.
}  

The rest of this  paper is organized as follows: First, in Sec.~\ref{sec:perturbation_theory} 
we investigate shortly the $T=0$ temperature properties
of the system in the gapped phases by means of perturbation theory. 
In Sec.~\ref{sec:semiclassical} we use the semiclassical approximation to obtain the 
low temperature correlation functions. Sec.~\ref{sec:QC} is devoted to the discussion 
of  the quantum critical regime, and our final conclusions are summarized 
in Sec.~\ref{sec:conclusions}. Some of the technical details have been relegated to 
Appendices.

\section{Perturbative analysis of $T=0$ temperature properties}
\label{sec:perturbation_theory} In this section we shall study the
scattering properties of quasiparticles in the $g\gg1$ and $g\ll
1$ regimes, where straightforward perturbation theory allows one
to obtain the energy of the quasiparticles, understand their
structure, and determine their scattering matrix.

\subsection{Paramagnetic side,  $g\gg1$}

We shall first study the limit of very large $g$. In this limit we
can perform an expansion in $1/g$ by taking the term describing
the effect of transverse field $\sim g$  as an unperturbed
Hamiltonian,
\begin{equation}
H_g=-jg\sum_{i} {\tilde{P}}_i \;,
\label{eq:H0_para}
\end{equation}
and considering the ferromagnetic interaction as a perturbation,
\begin{equation}
H_{\rm ferro}=- j\sum_{i} \sum_{\mu}
{\tilde{P}}^{\mu}_i  {\tilde{P}}^{\mu}_{i+1}\;.
\label{eq:H1_para}
\end{equation}
The ground state of Eq.~(\ref{eq:H0_para}) is simply
the paramagnetic state 
\begin{equation}
|\lambda_0 \rangle = \prod_i | \lambda_0 \rangle _i \; ,
\label{eq:para_alapall}
\end{equation}
where $|\lambda_0 \rangle _i$ denotes the state
$| \lambda_0 \rangle_i \leftrightarrow \frac{1}{\sqrt{Q}}(1,1,\dots,1)$ 
 at site $i$ in the basis of the
ferromagnetic eigenvectors.
Elementary excitations in the $g\to \infty$ limit consist of local 'spin flips'
of energy $\delta E= jg$,
where one of the spins is in a $\tilde{P}=0$ eigenstate  $|\lambda\rangle_i$, orthogonal to
$|\lambda_0 \rangle_i$. Obviously, there are $Q-1$ such states, and correspondingly,
the label $\lambda$ runs from $1$ to $Q-1$. We shall denote the state having  Potts spins 
flipped at sites $i,j,\dots$  by  $|i,\lambda_i;j,\lambda_j;\dots\rangle$.

Due to the ferromagnetic term, Eq.~(\ref{eq:H1_para}), these local excitations
can hop between lattice sites and 
get a dispersion. In leading order in $1/g$ the corresponding wave function
and quasiparticle energy of an elementary excitation is given by

\bea
|k ; \lambda \rangle &\approx& \sum_j e^{i x_j k}  |j,\lambda\rangle\;,
\\
\epsilon^{\lambda} (k) &\equiv& \tilde \epsilon(k)\approx jg(1- \frac{2}{Q} \frac{1}{g}  \cos{k})\;,
\eea
where $k$ is the quasi-momentum of the excitations, 
\remark
$x_j$ denotes the position of lattice site $j$, and we set the lattice constant to $a=1$.

It is also easy to compute the spin-spin correlation function in this approximation:
The operator $\tilde P^\mu$ creates  local spin flip excitations which then
propagate along the chain.  A simple calculation then gives  a single-particle
contribution to the dynamical susceptibility
\begin{equation}\label{eq:szuszcept_T=0_para}
\chi_{\mu\mu'}  (q,\omega) _{T=0} = \Bigl( \delta_{\mu\mu'} - \frac{1}{Q}\Bigr)
%\frac{A(g)}{\pi \Delta} 
%\frac{1}
\frac{A(g)\; 2\pi\; \Delta} 
{(\omega + i\delta)^2 - \epsilon_q^2 } + \dots \;,
\end{equation}
corresponding to a single particle contribution to the structure
function given by Eq.~(\ref{eq:structure_factor_para})  with
\be 
A(g) \Delta \sim \Delta^{4/15} \sim (g - g_c)^{2/9} 
\ee
as one approaches the quantum critical point.

Note that at $T=0$ temperature these
quasiparticles have an infinite lifetime to all orders in the perturbation theory 
(as guaranteed by energy and momentum conservation and the existence of a gap). 
Of course, Eq.~(\ref{eq:structure_factor_para}) only gives the leading
behavior of $S_{\mu\mu'}(\omega,q)$, and at higher energies a
many-particle continuum also appears above a threshold  due to
higher order corrections to the ground state wave function.

For  $g\gg 1$ the perturbation theory in $1/g$ is convergent.
Therefore the above picture holds to any order in $1/g$ for $g>
g_c$, apart from  the dispersion  $\tilde \epsilon(k)$ being
renormalized and the quasiparticle weight $A(g)$ being reduced for
finite values of $g$. In general the single particle contribution
to the correlation function reads,
\remark %corrected

\begin{equation}
S_{\mu,\mu'} (x,t) = \Bigl(\delta_{\mu,\mu'} - \frac1Q\Bigr)
 \int \frac{dk}{2 \pi} 
%\frac{A_k}{2 \epsilon_k}
\;A_k\;
 e^{i(kx-\tilde\epsilon(k) t)} \;,
\label{eq:korr_fv_T=0_Fourieralak_para}
\end{equation}
with the weight $A_k$ related to the matrix elements of $\tilde P^\mu$
between the exact quasiparticle state $|k ; \lambda\rangle$ and the ground 
state
$|0\rangle$ as $A_k = \frac{1}{Q-1}\sum_{\lambda,\mu} |\langle k ; \lambda| \tilde P^\mu_{x=0} |0\rangle|^2$. 

For large values of $x$ and $t$ the integral in Eq.~(\ref{eq:korr_fv_T=0_Fourieralak_para})
can be evaluated within the saddle point approximation, yielding 
\begin{equation}
S_{T=0}^{\mu,\mu'}(x,t) =
A(g)\; \left(\delta_{\mu,\mu'} - \frac{1}{Q} \right) \; 
K(x,t) \;,
%K_{\mu,\mu'}(x,t) \; R(\bar x ,\bar t) \;,
\label{eq:S_T=0_para}
\end{equation}
with $K(x,t)$ the Feynman propagator Eq.~(\ref{eq:Feynmanprop_para})
and $A(g) = \lim_{k\to0} \;A_k$.

The spin-spin correlation function further simplifies
in the vicinity of the quantum critical point
if we assume that the dispersion $\tilde \epsilon (k)$ takes on a relativistically
invariant form,
\be
\tilde \epsilon \approx \sqrt{\Delta^2 + c^2 k^2}\;,
\ee
with $c$ being the quasiparticle velocity,
and neglect the $k$-dependence of $A_k\to A(g)$.
In this case the zero temperature propagator can be expressed as
\remark %corrected
\begin{equation}\label{eq:Bessel_fuggv}
K (x,t) = 
\Delta\;
K_0 \left( \Delta \; \sqrt{x^2 - c^2 t^2} / c \right) ,
\end{equation}
where $K_0$ denotes the modified  Bessel function.
For $Q<4$ the quasiparticle gap $\Delta$ and the weight $A$ tend to zero
continuously as one approaches the critical values of the coupling,  $g_c$.
For $Q>4$, on the other hand, the phase transition remains first order, and
therefore both $A(g)$ and $\Delta$ remain finite at the transition point.

Having analyzed the single particle properties, let us now turn
our attention to two-particle states. We look for a two-particle
eigenstate of the Hamiltonian by making  the following ansatz in
leading order in $1/g$: 
\bea |k,k' \rangle & = &
\sum_{\lambda,\lambda'} \Bigl\{ A_{\lambda,\lambda'} \sum_{i<j}
\left( e^{i({k}{x_i} + {k'}{x_j})} |i,\lambda; j, \lambda' \rangle
\right)
\nonumber \\
& + & B_{\lambda,\lambda'} \sum_{i>j} \left( e^{i({k}{x_i} +
{k'}{x_j})} |i,\lambda; j, \lambda' \rangle  \right) \Bigr\}\;.
\label{eq:ansatz_2rhullamfg_para} \eea 

The coefficients
$A_{\lambda,\lambda'}$ and $B_{\lambda,\lambda'}$ in this ansatz
are related by the two-particle ${\cal S}$-matrix: 
\be
B_{\lambda,\lambda'} = \sum_{\tilde \lambda,\tilde \lambda'}
{\cal S}_{\lambda,\lambda'}^{\tilde \lambda,\tilde \lambda'}(k,k')
A_{\tilde \lambda,\tilde \lambda'} \;. 
\ee 

Substituting  (\ref{eq:ansatz_2rhullamfg_para}) 
into the Schr\"odinger equation we obtain an equation for
${\cal S}_{\lambda,\lambda'}^{\tilde \lambda,\tilde \lambda'}(k,k')$
that can be solved (see Appendix~\ref{Appendix} for details). The
expression of  ${\cal S}_{\lambda,\lambda'}^{\tilde \lambda,\tilde
\lambda'}(k,k')$ is rather involved for general values of $k$
and $k'$. However, rather remarkably, for $T \to 0$, where states with
both  $k$ and $k'$  going to zero are relevant, ${\cal S}_{\lambda,\lambda'}^{\tilde
\lambda,\tilde \lambda'}(k,k')$ reduces to the simple form
already given in the Introduction, Eq.~(\ref{eq:univ_s_matrix}):
\be
{\cal S}^{\lambda',\tilde \lambda'}_{\lambda,\tilde \lambda} \approx
(-1) \; \delta^{\lambda'}_{\tilde \lambda} \delta^{\tilde
\lambda'}_{\lambda} \label{eq:s_univ_para} \;. 
\ee 
Thus the scattering matrix assumes the very same  form as in
Ref.~[\onlinecite{Damle}]. While we obtained this equation in
first order perturbation theory, we believe that it is  valid to
all orders in  perturbation theory.

\subsection{Ferromagnetic side,  $g\ll1$}
\label{ferroside}

For $g\ll 1$ we can treat the ferromagnetic part of the Hamiltonian
(\ref{eq:H1_para}) as the unperturbed Hamiltonian, and consider the
'transverse field'  $g$, Eq.~(\ref{eq:H0_para})  as a perturbation.
The ground state of Eq.~(\ref{eq:H1_para}) is $Q$-fold degenerate and
corresponds to all spins being aligned in one of the $Q$ possible directions,
\be
|0\rangle_\mu  = \prod_j |\mu\rangle_j \;.
\ee
For $g=0$ the excitations of Eq.~(\ref{eq:H1_para}) consist of {\em domain walls},
\be
|\mu,\mu'\rangle_i  = \prod_{j\le i} |\mu\rangle_j \prod_{i< j'}
|\mu'\rangle_{j'} \;,
\label{eq:wall_mumu'}
\ee
and have energy $j$. The local field, Eq.~(\ref{eq:H0_para}), generates
a coherent motion of these domain walls. In leading order in $g$ the
wave function of the elementary excitations can be written as
\be
|k\rangle_{\mu,\mu'}  \approx \sum_j e^{i k x_j} |\mu,\mu'\rangle_j\;,
\ee
and their  energy is obtained by straightforward perturbation theory
as
\be
\epsilon_{k}^{\mu,\mu'}  = j \left( 1 - g \;\frac{2}{Q}
\cos{k} + \dots  \right)\;.\label{eq:ferro_dispersion}
\ee
Thus quasiparticles have a gap $\Delta(g)\approx j( 1 - 2g / {Q})$
in this phase too, and the ground state is stable.

In the ground states the expectation value of the operators
$ \tilde P^\mu$ is finite,
\be
\langle \tilde \mu| \tilde P^\mu | \tilde \mu \rangle =
\left\{
\begin{array}{cl}
m \;(Q-1)/Q\;, & $ if $ \mu = \tilde \mu ,\cr
-m/Q \;, & $ if $ \mu \ne \tilde \mu.
\end{array}
\right.
\label{eq:<tildeP^mu>}
\ee

Correspondingly, the structure  factor  has a delta peak associated with
the long range order at $\omega = q=0$, given by Eq.~(\ref{eq:struktfakt_ferro_alapall}).
We remark here that any finite temperature induces a
dilute gas  of domain walls which destroy this long range order
and broaden the delta peak. As we show in the next section,
this delta peak has a Gaussian broadening for $Q=2$, however, becomes a
diffusive pole for any $Q>2$.

Similar to the paramagnetic phase discussed in the previous subsection,
all the above perturbative results carry over to any finite $g<g_c$,
because perturbation theory is convergent. For $Q<4$ the order parameter
vanishes as one approaches $g_c$,
\remark %check this
\be
m(g) \sim (g_c - g)^\beta\;,
\ee
with $\beta$ the order parameter exponent for
the two-dimensional classical $Q=3$ state Potts model ($\beta_3 = 1/9 $)
%G checked exponents
and the two-dimensional Ising model ($\beta_2 = 1/8 $), respectively,\cite{DiFrancesco} 
and the quasiparticle gap vanishes with the same exponent as in the
paramagnetic phase, $\nu = 5/6$ for $Q=3$ and $\nu=1$ for $Q=2$.

We now turn to the  study two-particle properties in the limit
$g\ll1$. To keep track of domain wall excitations, it is worth
labelling multi-domain wall configurations in a slightly different
way as before. Suppose that we have a domain wall with a
polarization  $ \mu $ on its left and  $ \mu' $ on the right. We
can then define the quantum number $\theta=1,\dots,Q-1$ of this
domain wall as the size of the step between the two sides, \be
\mu'= (\mu + \theta) \;{\rm mod}\; Q\;. \ee Clearly, to
characterize any configuration, it is sufficient to give the
vacuum state on the left of the chain, and then specify the
quantum numbers $\{\theta_j\}$.  With this notation, we can thus
denote the state described by Eq.~(\ref{eq:wall_mumu'}) as
\begin{equation}
\label{eq:hullamfv_def_ferro}
| \mu , \mu' \rangle_i \equiv  | i, \theta \rangle_\mu \;.
\end{equation}

With the above notation, the two-particle wave function has the
following form in leading order in $g$,
\bea
|k, k' \rangle & \equiv &
\sum_{\theta,\theta'}
\Big\{ A_{\theta,\theta'} \sum_{i<j} \left( e^{i({k}{x_i} +
{k'}{x_j})}
|i, \theta; j,\theta' \rangle_\mu  \right) +
\nonumber
\\
&& B_{\theta,\theta'} \sum_{i>j} \left( e^{i({k}{x_i} +
{k'}{x_j})} | i,\theta; j, \theta' \rangle_\mu  \right) \Big\} \;.
\label{eq:2p-hullamfg_ferro} \eea 
Again, a simple calculation
outlined in Appendix~\ref{Appendix} provides us the scattering
matrix ${\cal S}_{\tilde \theta, \tilde \theta'}^{\theta,\theta'}$
relating the amplitudes  $A_{\theta,\theta'}$ and
$B_{\theta,\theta'}$ of the incoming and outgoing particles,
respectively. In the limit of vanishing  momenta, $k,k'\to0$  we
find 
\be {\cal S}_{\tilde \theta, \tilde \theta'}^{\theta,\theta'} =
(-1)\; \delta_{\tilde \theta}^{\theta'} \delta_{\tilde
\theta'}^{\theta} \;, \label{eq:s_univ_ferro} \ee 
{\em i.e.} quasiparticles  scatter as 'hard balls'. This equation can also be
visualized as a condition that the orientation of the order
parameter between two colliding domain walls with vanishing
quasi-momenta remains unchanged after the collision. Again, we
believe that the exchange form of the scattering matrix,
Eq.~(\ref{eq:s_univ_ferro}), remains valid to all orders in $g$
for $g<g_c$.

\begin{figure}[tb]
\centering
\includegraphics[width=5cm]{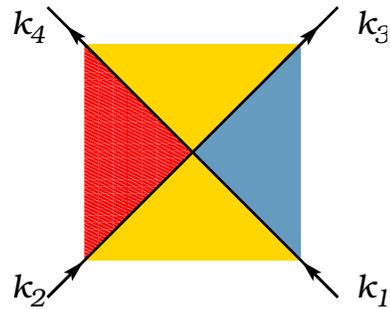}
\caption{\label{abra:ferro_smatrix} (Color online) Sketch of colliding domain walls on
the ferromagnetic side at $T\ll \Delta$ as $k \to 0$. Note that the 
polarization of the middle domain must be conserved and so the order of
 the domains along the chain does not change with time. }
\end{figure}

\section{Semiclassical approximation in the gapped phases}

\label{sec:semiclassical}

In this section, we shall study dynamical correlation functions
in the $T\ll\Delta$ limit in the gapped phases,
following the  semiclassical treatment of Refs.~[\onlinecite{sachdev_young,Damle,tagging}].
This approach is based on the observation that at very low temperatures
only quasiparticles with energy close to the quasiparticle
gap are present, $\epsilon \approx \Delta$.
The energy of these quasiparticles can be approximated
as
\begin{equation}\label{eq:k_to_0_approx_ferro}
\epsilon_k = \Delta + \frac{c^2}{\Delta} \; \frac{k^2}{2} +
{\cal O}(k^4)\;,
\end{equation}
with $c$ a constant playing the role of the speed of light,
and $\Delta/c^2$ the mass of the quasiparticles.
For $T \ll \Delta$ the distribution of quasiparticles
is described simply by Boltzmann statistics,
\begin{equation}\label{eq:klasszikus_n(k)_ferro}
n(k) \sim  e^{-\beta \Delta} \; e^{-\beta \; \frac{c^2 k^2}{2
\Delta}} \;,
\end{equation}
with $\beta=1/T$ the inverse temperature. Correspondingly, the
quasiparticle density is exponentially small
\begin{equation}\label{eq:suruseg_ferro}
\varrho  = (Q-1) \; \sqrt{\frac{T\Delta}{2 \pi c^2}} \;
e^{-\Delta/T} \;,
\end{equation}
and the typical separation between them,
$d_T \sim 1/\varrho$ increases exponentially at low temperatures,
$d_T \sim e^{\Delta/T}$. This must be compared
to the De Broglie wavelength of the particles, $\lambda_T$, giving the quantummechanical
extension of the quasiparticles' wave function. This latter is
given by the inverse of the typical  momentum $k$ of the quasiparticles,
$\lambda_T \sim c/\sqrt{T\Delta}$.
Clearly, at very low temperatures  the average separation of the quasiparticles
is much bigger than their quantum mechanical size,
\begin{equation}\label{eq:semiclass_requi_ferro}
d_T \gg \lambda_T \;,
\end{equation}
which makes one  possible to treat the quantum mechanical state of the
system in both phases within the semiclassical approximation
for $T\ll \Delta$.

Unfortunately, in one dimension  neighboring particles cannot
avoid collisions with each other, thus they will get unavoidably
closer than the De Broglie wavelength, where quantum mechanics is
at work. These collisions must therefore be described within the
framework of quantum mechanics. Fortunately, at low $T$, where the
semiclassical limit is valid, the system is dilute enough so that
we have only  2-particle scattering. Furthermore, the colliding
quasiparticles have momenta $k\sim 1/d_T$. Therefore, at low
enough temperatures, we can use the simple forms of the
scattering matrices given by  Eqs.~(\ref{eq:s_univ_para}) and
(\ref{eq:s_univ_ferro}). This simple form of the ${\cal S}$-matrix will
enable us
to obtain analytical results for the correlation functions
within the semiclassical picture.

\subsection{Semiclassical correlation function on the ferromagnetic side}

Let us first compute the spin-spin correlation in the less complicated case of
ferromagnetic  phase.
By definition, the time-dependent correlation function is defined as
\begin{equation}
S_{\mu,\mu'}(x,t) = \langle
e^{itH} {\tilde P}^{\mu}(x) \; e^{-itH}
\; {\tilde P}^{\mu'}(0)  \rangle
 \;,
\label{eq:korrfv_T=0_ferro}
\end{equation}
where $\langle\dots \rangle$ denotes now thermal averaging over all possible
many-body states,
\be
\langle\dots \rangle = \sum_n \langle n|\dots |n\rangle e^{-\beta E_n}\;,
\label{eq:aver_mb}
\ee
where the summation runs over all many-body eigenstates $ |n\rangle $
of energy $E_n$ of the total Hamiltonian.
Within the semiclassical approximation we can replace this
average by an average over all possible quasiparticle configurations,
{\em i.e.}, by an average over the quasiparticle velocities $v_\nu$,
their positions $x_\nu$, and internal quantum numbers $\theta_\nu$,
($\nu=1,\dots,M$ with $M$ the number of quasiparticles),
\bea
\langle\dots \rangle &\approx&
\sum_{\{\theta_\nu\}}
\int \prod_\nu dx_\nu \prod_\nu dv_\nu
P (\{ x_{\nu},v_{\nu},\theta_{\nu} \})
\nonumber \\
&&\langle  \{x_{\nu},v_{\nu},\theta_{\nu} \}|\dots
|\{x_{\nu},v_{\nu},\theta_{\nu} \}\rangle \;,
\eea
where the distribution function $P (\{ x_{\nu},v_{\nu},\theta_{\nu} \})$ is
simply
\bea
P (\{ x_{\nu},v_{\nu},\theta_{\nu} \}) &=& \frac{1}{L_x^M} \;
\frac{1}{(Q-1)^M}  \prod_{\nu} P(v_\mu)\;,
\label{eq:termikus_suly}
\eea
with  $L$ the system size and $P(v_\mu)$ the Boltzmann distribution
of the quasiparticle velocities $v= c^2 k/\Delta$,
\begin{equation}\label{eq:Boltzmannfaktor}
P {(v)=\sqrt{\frac{\Delta}{2\pi c^2 T}} \;
e^{-\frac{\Delta}{c^2 T} \; \frac{v^2}{2}}} \;.
\end{equation}

Of course, in the above equations an average over $M$ should be
taken. In the $L\to \infty$ limit, however, we can replace  $M$ by
the average particle number, $M\to \varrho \;L$, without changing
the final result.

\begin{figure}[h]
\centering
\includegraphics[width=8cm]{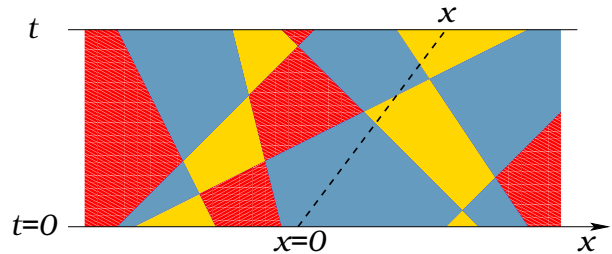}
\caption{\label{abra:korrelaciosfuggv_ferro} (Color online) Time evolution 
of domain walls  on the  ferromagnetic side at $T \ll \Delta$. For the correlation function
 $S_{\mu \mu'}(x,t)$ we need to take the average over all domain configurations. }
\end{figure}

To evaluate the correlation function $S_{\mu \mu'}(x,t)$, let us
first compute the probability $Q_{\mu \mu'}(x,t)$,  that at time
$t=0$  and position $x=0$ the order parameter points into
direction $\mu$ while  at time $t$ and position $x$ it points into
direction $\mu'$. To evaluate this probability, we notice that the
number of domain walls is conserved in course of every domain wall
collision. Furthermore, the 'color' (orientation) of the $q$'th
domain from the left remains unchanged due to the exchange form
of the scattering matrix (see
Fig.~\ref{abra:korrelaciosfuggv_ferro}). As a consequence, $Q_{\mu
\mu'}(x,t)$ can be simply written as 
\bea 
&&Q_{\mu,\mu'}(x,t)  = \sum_{l,k} P((0,0)\; \mbox{in $l$'th}) \nonumber 
\\
&&\phantom{nn} P((x,t)\; \mbox{in $k$'th})\; P(l,\mu; k,\mu')\;,
\label{eq:korrfv_nnel_ferro} 
\eea 
{\em i.e.} it is
simply a product of the probability $ P((0,0)\; \mbox{in $l$'th})$
that $(t=0,x=0)$  is within the $l$'th domain from the left, while
the $(t,x)$ point lies in the $k$'th domain, and probability that
the $l$'th domain points in direction $\mu$ while the $k$'th in
direction $\mu'$, $ P(l,\mu; k,\mu')$. We have to remark here that
the orientation of domain walls is {\em correlated}. This
correlation is simply generated by the constraint that if a domain
points in a direction $\mu$ then its neighbor must point into one
of the {\em other} $Q-1$ directions. This correlation is hidden in
$ P(l,\mu; k,\mu')$, which can be computed by constructing a
Markov equation, and depends only on $n\equiv l-k$, $P(l,\mu;
k,\mu') = P(n, \mu,\mu')$:
\bea
P(n, \mu,\mu') = \frac{1}{Q} \left(
{\frac{-1}{Q-1}}\right)^{|n|} \Bigl(  \delta_{\mu,\mu'} - \frac1Q
\Bigr) +\frac1{Q^2} \;. %\nonumber 
\label{eq:Rmatrixelem_ferro}
\eea
 Note that this correlation function decays exponentially for
any $Q>2$, while it oscillates for the Ising model, $Q=2$.

The first part of Eq.~(\ref{eq:korrfv_nnel_ferro}) also
depends only on the separation $n=l-k$ and the coordinates
$x$ and $t$. Therefore, Eq.~(\ref{eq:korrfv_nnel_ferro})
can also be rewritten as
\be
Q_{\mu,\mu'}(x,t)  =
\sum_{n=-\infty}^\infty  D(n,\;(x,t))
\;P(n, \mu,\mu')\;,
\label{D:sum_n}
\ee
with $P(n, \mu,\mu')$ given by Eq.~(\ref{eq:Rmatrixelem_ferro}) and
\be
D(n,\;(x,t)) \equiv
\sum_{l} P((0,0)\; \mbox{in $l$'th}) P((x,t)\; \mbox{in $l+n$'th})\;.
\ee
Clearly, $D(n,\;(x,t))$
is just the probability that the domain of $(x,t)$
is the $n$'th domain to the right from the domain of $(0,0)$.

The probability $D(n,\;(x,t))$ can be computed as follows.
First, following Refs.~[\onlinecite{Damle,tagging}], let us introduce
the notion of 'particles'.
In the configuration space, domain walls are located along straight lines $\nu$,
\begin{equation}\label{eq:trajektoria_egyenlete_para}
x_\nu (t) = x_\nu + v_\nu \, t \;,
\end{equation}
where  $x_\nu$ and $v_\nu$ are the position and velocity of the
$\nu$'th domain wall  at $t=0$. 'Particles', however, correspond
to a given step $\theta$ and are 'reflected' when two lines cross.
There is a simple way to tell which particle $p$ is moving along
line $\nu$ at time $t$. This is given by the function $p_\nu(t)$
which we chose to coincide at $t=0$ with the index  $\nu$ of a
line, $p_\nu(t=0)=\nu$. Particles are, however, impenetrable.
Therefore if line $\nu$ is crossed by another line from the left then 
$p_\nu$ decreases by one, while if it is crossed by a line from the right,
then it increases by one. 
Therefore, to keep track of   $p_\nu(t)$ we just have to count the
number of lines that crossed $x_\nu(t)$ from the left and from the
right since time $t=0$,
\begin{equation}\label{eq:traj_sorszam_t'ben_para}
p_\nu(t) = \nu  + \sum_{\nu'} \left\lbrack \Theta (x_\nu (t)-x_{\nu'} (t))
- \Theta (x_\nu -x_{\nu'}) \right\rbrack \;,
\end{equation}
where $\Theta$  denotes  the step function.

Similarly, 
for a {\em fixed}  set of lines, $\{ x_{\nu},v_{\nu} \}$,
the probability that the straight line from $(0,0)$ to $(x,t)$ moves across
$n$  domains to the right is simply
\begin{equation}
\label{eq:P_(x,t)_n-nel_jobbra}
P(n |\{ x_{\nu},v_{\nu} \} , (x,t)) = \delta_{n, \sum_{\nu}
\left\lbrack \Theta (x-x_\nu (t)) - \Theta (-x_\nu) \right\rbrack} \;,
\end{equation}
To obtain $D(n,(x,t))$, we should average this expression over
$\{ x_{\nu},v_{\nu} \}$. To do this we just introduce the
following integral representation of the Kronecker-delta,
\bea
\label{eq:Diracdelta_integrallal}
&&\delta_{n, \sum_\nu \left\lbrack \Theta (x-x_\nu (t)) - \Theta (-x_\nu)
\right\rbrack} =
\nonumber \\
&&\phantom{nnn} \int_{-\pi}^{\pi} \frac{d \phi}{2\pi} e^{i \phi
\; \left( n - \sum_\nu \left\lbrack \Theta (x-x_\nu (t)) -
\Theta (-x_\nu ) \right\rbrack \right) } \;.
\eea 
The advantage of
this form is that the averaging factorizes and can be evaluated
analytically. With some algebra we obtain that 
\bea &&D(n,(x,t)) =
\int_{-\pi}^{\pi} \frac{d \phi}{2\pi} e^{i n\phi } I(x,t)^M \;,
\label{D}
\\
&&
I(x,t) =
\langle e^{-i \phi \{\Theta (x-x_\nu - v_\nu t ) -
\Theta (-x_\nu) \}}\rangle_{v_\nu,x_\nu} \;.
\eea
The integral $I(x,t)$ can then be simply evaluated to yield
\begin{equation}\label{eq:reszintegral_eredmeny}
I(x,t) = 1 - \frac{\bar t}{\bar {L _x}} \left\{ G(u)+G(-u) - e^{i
\phi} G(u) - e^{- i \phi} G(-u) \right\},
\end{equation}
where
\begin{equation}
\label{eq:G_fuggveny_def}
G(u) = \frac{1}{2 \sqrt{\pi}} e^{-u^2} - \frac u2 \; {\rm erfc}(u);
\end{equation}
and we introduced dimensionless time and length,
$\bar x = x / \xi_c$, and $\bar t = t / \tau$ and the corresponding dimensionless
velocity $ u = \bar x / \bar t$,
with $\xi_c$ and $\tau$ the characteristic  classical
correlation length and time
\bea
\xi_c & \equiv& 1/ \varrho = \frac{1}{Q-1} \; \sqrt{\frac{2\pi
c^2}{T\Delta}} e^{\Delta/T},
\label{eq:xi_c_def}
\\
\tau &\equiv& \frac{1}{Q-1} \; \frac{\sqrt{\pi}}{T} e^{\Delta/T}\;.
\label{eq:tau}
\eea
Finally, $I(x,t)^M$ in Eq.~(\ref{D})  can be re-exponentiated as
\be
I(x,t)^M = e^{- {\bar t} \left\{ G(u)+G(-u) - e^{i
\phi} G(u) - e^{- i \phi} G(-u) \right\}}.
\ee
Substituting Eqs.~(\ref{D}) and
(\ref{eq:Rmatrixelem_ferro}) into Eq.~(\ref{D:sum_n})
the sum over $n$ can be computed using the identity
\begin{equation}\label{eq:szum_n_elvegez}
\sum_{n= - \infty}^{\infty} e^{i \phi \; n} \; \left(
\frac{-1}{Q-1} \right) ^{|n|}  = \frac{(Q-1)^2-1}{
(Q-1)^2+1+2(Q-1)\cos{\phi}  },
\end{equation}
and the final form of the
function $Q_{\mu\mu'}(x,t)$  reads:
\be
Q_{\mu\mu'}(x,t)
 = \frac1{Q^2} +
R(\bar{x},\bar{t})
\frac{1}{Q} \Bigl(  \delta_{\mu,\mu'} - \frac1Q \Bigr)
\;,
\ee
with the relaxation function given by Eq.~(\ref{eq:relax_fv_integralalak}).
The correlation function $S_{\mu\mu'}(x,t)$ can then be easily computed
in terms of $Q_{\mu\mu'}(x,t)$ using  Eq.~(\ref{eq:<tildeP^mu>}),
as
\be
S_{\mu\mu'}(x,t) = \sum_{\tilde \mu, \tilde\mu'}
\langle \tilde P^\mu\rangle_{\tilde \mu} \langle \tilde P^{\mu'}\rangle_{\tilde \mu'}
\;Q_{\tilde \mu,\tilde \mu'}(x,t)\;.
\ee
Putting all this together we finally obtain
\be
S_{\mu\mu'}(x,t)
 = \frac{m^2}{Q}
\Bigl(  \delta_{\mu,\mu'} - \frac1Q \Bigr)\;R(\bar{x},\bar{t})
\;.
\ee

\begin{figure}[h]
\centering
\includegraphics[clip,width=7cm]{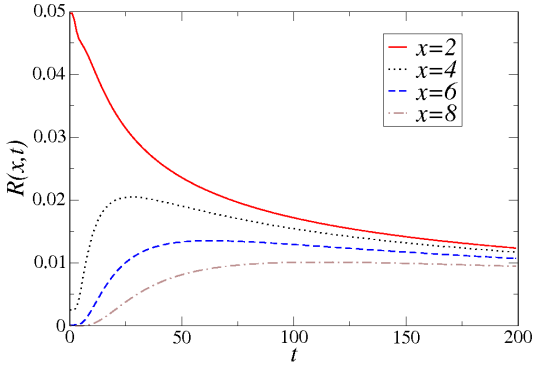}
\caption{\label{abra:relaxfuggv} (Color online) The relaxation function 
$R(\bar{x},\bar{t})$ as a function of $\bar{t}$ for different 
values of $\bar{x}$ at $Q=3$.
}
\end{figure}

The time evolution of $R(\bar{x},\bar{t})$ is shown in Fig.\ref{abra:relaxfuggv}
The behavior of the relaxation function $R(\bar x, \bar t)$ is quite interesting for  $\bar t \gg 1$.
The integrand in the definition (\ref{eq:relax_fv_integralalak}) goes to zero 
exponentially as $\bar t \to \infty$, and only a small range 
around 
$\phi \approx 0$ contributes to the integral.
%the minimum of the integrand, which is at 
Hence the relaxation function can be approximated as
\be
R(\bar x,\; \bar t \gg 1) \approx \int_{-\infty}^{\infty} {d\phi\over 2\pi\; Q} \; 
(Q-2)
\; e^{-\bar t\; F(u) \phi^2/2}
\; {\rm cos}(\phi \bar x) \;, 
\label{eq:relaxfv_diffuziv_1}
\ee
where $F(u)=1/\sqrt{\pi}e^{-u^2}+u \; {\rm{erf}} (u)$. 

This integral can be calculated analytically and is given by 
\be
R(\bar x, \bar t \gg 1) \approx 
\frac{Q-2}{ Q}
\sqrt{\frac{1}{2\pi\;\bar t \;F(u)}} e^{-\frac{ \bar{x}^2}{2 \bar{t} F(u)}} \;.
\label{eq:relaxfv_diffuziv_2}
\ee
For
 $\bar t \gg \bar x$ we have $F(u) \approx 1/\sqrt{\pi}$, 
and the relaxation function assumes a diffusive form
\be
R(\bar x \ll \bar t) \propto {\xi_c\over\sqrt{4 \pi t D }}\; e^{-x^2/4Dt} \;,
\label{eq:relaxfv_diffuziv_3}
\ee
with the diffusion constant defined as 
\be 
D=\frac{1}{2\sqrt{\pi}} {\xi_c^2\over\tau} \;.
\label{eq:diffus_const}
\ee
As a result, the dynamical structure factor has a diffusive structure at 
$\omega,q \to 0$, as already discussed in the Introduction.

\subsection{Semiclassical dynamics in the quantum paramagnetic phase}

The analysis of the previous subsection can also be extended to  the paramagnetic
phase.  Similar to the ferromagnetic phase, the average
over the many-body eigenstates in Eq.~(\ref{eq:aver_mb}) can be replaced in the semiclassical limit
by an average over all possible initial states,
 $|\{x_{\nu},v_{\nu},\lambda_{\nu} \}\rangle \;$.
The main difference in the calculation is that on the paramagnetic
side the operator $\tilde P^\mu(0)$ in
Eq.~(\ref{eq:korrfv_T=0_ferro})  creates a  quasiparticle with
some internal quantum number $\lambda_0$, and velocity  $v$  at
time $t=0$ with some probability amplitude $e^\mu_{\lambda_0}(v)$
in addition to the already existing quasiparticles. This particle,
together with the other quasiparticles, propagates under the
action of $H$ in Eq.~(\ref{eq:korrfv_T=0_ferro})) and collides
with them. For very small temperatures, however, the two-particle
scattering matrix takes on an exchange form, and therefore
particles only exchange their velocity while conserving their
internal quantum numbers.

\begin{figure}[h]
\centering
\includegraphics[width=8cm]{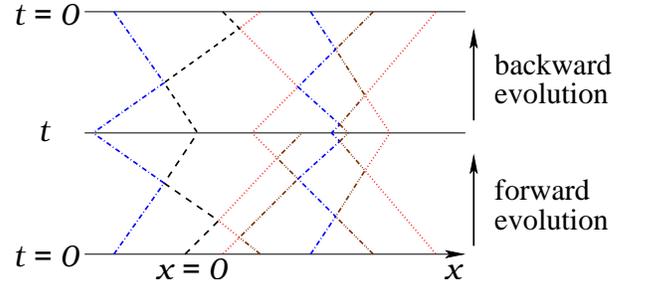}
\caption{\label{abra:idofejlesztes_para} (Color online)
To evaluate the correlation function on the paramagnetic side, one must 
keep track of   the particle created by the operator
$\tilde P^{\mu}$. Lines drown with different line-styles denote trajectories 
of different 'particles'.
The two states after and before  the  forward and backward time evolution must be  
identical. Therefore, the internal quantum numbers of the particles must 
obey constraints.}
\end{figure}
As a consequence, at any time precisely one of the particles will
have the velocity $v$ of the initial particle and it will be at
position $v t$. It is easy to see that this very particle must be
annihilated at
 time $t$ by $\tilde P^{\mu'}(x)$
otherwise the final state obtained after the action of $e^{itH}$
in Eq.~(\ref{eq:korrfv_T=0_ferro}) will be orthogonal to
the  initial state.   (The evolution of these quasiparticles is
shown in Fig.~\ref{abra:idofejlesztes_para}.) The probability
amplitude that this particle is annihilated is proportional to
$(e^\mu_{\lambda'}(v))^* e^{ikx}$ with $k = \Delta  \; v/c^2$ the
quasi-momentum that corresponds to $v$ and $\lambda$ the internal
quantum number of the particle that is removed. As we shall see
below,  $\lambda'$ in this expression must be equal to
$\lambda_0$. This shall be guaranteed by  another requirement,
namely that the internal quantum numbers in the final state must
be exactly the same as those of the initial state. To see what
this condition implies let us assume that   in the initial state
we had quasiparticles with quantum numbers
$\{\lambda_1,\lambda_2,\dots\}$ at positions $x_\mu$
($x_1<x_2<\dots$) with velocities $v_\mu$, and that  the new
particle is created right after the $p$'th particle 
\bea 
\tilde P^\mu(0) : &\;\{\dots,\lambda_p,\lambda_{p+1},\dots\} \\
&\longmapsto \{\dots,\lambda_p,\lambda_0,\lambda_{p+1},\dots\}\;. \nonumber
\eea
Let us now define the function $p_0(t)$ which gives us the label of that
particle which has velocity $v$ at time $t$. Obviously,
$p_0(t=0)=0$. 
%\textbf{?ez igy nem teljesen konzisztens az elozo keplettel:nem p kene??}.
The important observation is that the above {\em order} of these quantum numbers from the left to the right along the chain  does not change under the collisions due to the simple
form of the scattering matrix Eq.~(\ref{eq:s_univ_para}). Suppose now that
after the time evolution $t$ particle '$n+p$' moves with velocity $v$, {\em i.e.}, 
$p_0(t)$ takes the value $p_0(t) = p+n$. (Here we assume for the sake of simplicity that
$n>0$, but the derivation can be easily generalized for $n\le0$.) 
Since  the operator $\tilde P^{\mu'}$ must destroy
precisely this particle of velocity $v$, $\tilde P^{\mu'}$
therefore changes the series of quantum numbers as 
\bea
\tilde P^{\mu'} : \; &\{\dots,\lambda_p,\lambda_0,\lambda_{p+1},\dots \\
&\dots,\lambda_{p+n-1},\lambda_{p+n},\lambda_{p+n+1},\dots\}  \nonumber \\
&\longmapsto \{\dots,\lambda_p,\lambda_0,\lambda_{p+1},\dots \nonumber \\
&\dots, \lambda_{p+n-1}, \lambda_{p+n+1},\dots \}\;. \nonumber
\eea 
Since the order of these quantum numbers does not change under the
action of $e^{-itH}$ either, requiring that the internal quantum
numbers of the initial states be the same as those of the final
state amounts in the condition 
\be \lambda_0 \equiv \lambda_{p+1}
\equiv \dots \equiv \lambda_{p+n}\;. \label{eq:lambda_condition}
\ee 
Now, similar to the ferromagnetic case, the value of $n$ in
the previous expressions can be simply computed as the number
$N_+$ of lines  $x_\nu(t)$ that cross the line connecting $(0,0)$
with $(x,t)$ from the right minus the number of lines $N_-$ that
cross it from the left, 
\be n = \sum_\nu [\Theta(x(t) -x_\nu(t)) -
\Theta(0 -x_\nu)] \;, 
\ee
 where $\nu$ labels the quasiparticles in
the initial state, $x_\nu(t) = x_\nu + v_\nu t$, and $x(t) = v t$
is the trajectory of the new particle with velocity $v$ 
is  created by $\tilde P^\mu(0)$ at the origin, $x(0)=0$.

To compute the contribution of this particular state to the
correlation function we must consider the important  detail of
phase factors. Quasiparticles in the initial state generate a
phase factor  $e^{-i t \sum_\nu \epsilon(v_\nu)}$ under the action
of $e^{-itH}$. This is, however, completely canceled under the
action of $e^{itH}$, except for the quasiparticle created by
$\tilde P^\mu$ giving a factor  $e^{-i t \epsilon(v)}$.
Furthermore, every collision results in a sign change of the
many-body wave function. All these signs cancel under the forward
and backward propagation, excepting the ones that are associated
with collisions with the extra particle. These give an extra sign
$(-1)^{N_+ + N_-}$, which can, however be also written  more
conveniently as 
\be 
S= (-1)^{N_+-N_-} = (-1)^n\;. 
\ee 

Putting all
these together we obtain the following expression for the
correlation function,
\begin{widetext}
\be
S_{\mu\mu'}(x,t) =
\left(\sum_\lambda \int dv\; (e^\mu_{\lambda}(v))^* e^{\mu'}_{\lambda}(v)\;
 e^{-i(k(v) x-\epsilon(v) t) } \right)  \;
\Bigl\langle \sum_{n=-\infty}^\infty {(-1)^n \over (Q-1)^{|n|}}
\delta_{n,\sum_\nu [\Theta(x -x_\nu(t)) -  \Theta(0 -x_\nu)]}
\Bigr\rangle_{\{v_\nu,x_\nu\}}\;,
\ee
\end{widetext}
where the factor $1/(Q-1)^{|n|}$ comes from the condition
Eq.~(\ref{eq:lambda_condition}) after averaging over all possible
internal quantum numbers. The first term in this expression gives just the $T=0$
correlation function Eq.~(\ref{eq:S_T=0_para}), while  the second part is  the 
same relaxation
function as the one found in the ferromagnetic phase, 
Eq.~(\ref{eq:relax_fv_integralalak}), 
\begin{equation}
S_{\mu,\mu'}(x,t) =S_{T=0}^{\mu,\mu'}(x,t) \; R(\bar x, \bar t)\;.
\label{eq:korrfv_vegsoalak_para}
\end{equation}

Finally, let us briefly discuss the properties of quasiparticle density-density 
correlations in the paramagnetic phase. The density of quasiparticles 
with  quantum number $\lambda$ is simply given by 
\be
\varrho_\lambda (x,t)  = \sum_\nu \delta_{\lambda,\lambda_{p_\nu(t)}} \delta(x_\nu(t) - x)\;,
\ee
where the summation runs over all lines. The average density,
$\varrho \equiv \langle \varrho(x,t)\rangle $ is simply given by 
Eq.~(\ref{eq:suruseg_ferro}). The density-density correlation function 
can be computed analytically within the semiclassical approximation. We shall 
not discuss this calculation here, since it is basically identical 
to the computation of Damle and Sachdev.\cite{Damle} The result is simply 
\begin{widetext}
\be
\langle \varrho_\lambda (x,t)\varrho_{\lambda'} (0)\rangle 
= \varrho^2 \;\frac{\delta_{\lambda,\lambda'}}{Q-1}\;
e^{-\bar t\;(G_+ + G_-)}
\times \left\{ 
\left(
\frac{e^{-u^2}}{\sqrt{\pi} \; \bar t} + 2\; C_+\;C_- 
\right) I_0(2 \bar t\sqrt{G_+ G_-}) 
+
\frac {C_+^2 \; G_- + C_-^2 \; G_+} {\sqrt{G_+ G_-}}  \;
I_1(2 \bar t\sqrt{G_+ G_-}) 
\right\}\;,
\ee
\end{widetext}
%\narrowtext
where $G_\pm$ and $C_\pm$   denote the functions 
\bea 
G_\pm \equiv G(\pm u) & = & {1\over 2\sqrt{\pi}} \; e^{-u^2} \mp  \frac u 2 \; {\rm erfc}(\pm u)\;, 
\\
C_\pm  \equiv C(\pm u)  & = &  \frac1  2 \;{\rm erfc}(\pm u)\;, 
\eea
with $u = \bar x/\bar t$ the dimensionless velocity defined earlier.
%and the dimensionless quantities defined as 
%$\bar x = \varrho \; x$, $\bar t = \sqrt{2T/\Delta} \varrho$.
%and $I_{0,1}$ denote the ???.

For large time scales, $\bar t \gg 1$ the density-density correlation 
function also displays a diffusive behavior,\cite{Damle}
\be 
\langle \varrho_\lambda (x,t)\varrho_{\lambda'} (0)\rangle \sim 
\delta_{\lambda,\lambda'} \; {1\over\sqrt {\bar t}} \; e^{-\sqrt{\pi} \bar x^2 /2 \bar t^2}\;,
\ee
in agreement with the  behavior of the relaxation function $R(\bar x, \bar t)$.
In the $Z_3$ case,  this correlation function also bears physical meaning and 
corresponds to the correlation function of the {\em chirality density}.

\section{Dynamics in the quantum critical region}

\label{sec:QC}

In this section we shall focus to the 
$Q=3$ Potts model which has a quantum critical point  at $T=0$ and $g=g_c$.
%\akos{ We cannot construct the quasiparticles in the quantum critical region of the $Q=3$ quantum Potts model, and thus we have to choose a different approach to investigate the dynamics in this region, than we used in the gapped phases.}
We can gain many information already  from the quantum-classical mapping, by just  
using the scaling properties of the singular part of the free energy density:
\be 
f(g-g_c, h , T) = b^{-2} \; f(b^{y_t}(g-g_c),b^{y_h}  h , T b )\;,
\label{eq:scaling}
\ee 
where $y_t = 2 - x_t$ and $y_h = 2 - x_h$ denote the scaling dimensions
of the temperature and the magnetic field in the classical Potts model, 
and $x_t = 4/5$ and $x_h = 2/15$ denote the dimensions of the corresponding 
primary fields that are  known from conformal field theory.\cite{DiFrancesco,Potts_CFT} 
Note that the temperature plays the role of a finite 
system size while the coupling $g$ corresponds to the temperature. 

From Eq.~(\ref{eq:scaling}) immediately follows that the gap vanishes as
$
\Delta \sim |g-g_c|^{5/6}\;, 
$ 
and that the susceptibility behaves in the quantum critical region as
%diverges at the quantum critical point as
\be 
\chi(T>\Delta) \sim \frac 1 {T^{26/15}}\;.
\ee

For $T< \Delta$ the susceptibility becomes finite, but it diverges as one approaches the
quantum critical point, 
\be 
\chi(T=0) \sim \frac 1 {|g-g_c|^{13/9}}\;.
\ee

So far, we only discussed thermodynamical properties.
However, a lot of information  can also obtained by making use of the conformal invariance
of the critical theory. At the critical point, the imaginary time correlation function is scale 
invariant and has  the following form:\remark
\bea
\chi_{\mu \mu'}^{T=0} (x,\tau) &=& (\delta_{\mu \mu'} -\frac1Q) \;\chi(x,\tau)\;,
\\
\chi(x,\tau) &\sim&  \frac{1}{(\tau^2 + x^2)^{x_h}}.
\label{eq:corrfunc_critpoint}
\eea 
For simplicity,  we set the 'speed of light' to one in this section.

To obtain the {\em finite temperature} imaginary time susceptibility
(retarded correlation  function),
we  introduce new, complex coordinates $z \equiv \tau + i \; x$ and 
$\bar z = \tau - i \; x$ and rewrite (\ref{eq:corrfunc_critpoint}) as
\be
\chi^{T=0} (z,\bar z) \sim 
\frac{1}{z^{x_h} \;\bar z^{x_h}}.
\label{eq:corrfunc_critpoint_2}
\ee 
Then we map the complex plane to a strip of width $\beta = 1/T$ by the 
transformation
\be
w =\frac {1}{\pi \; T} \; {\rm cotan} \left( \pi \; T \; z \right) \;, 
\label{eq:tangens_mapping}
\ee
and use the transformation properties of primary fields to obtain\cite{DiFrancesco} 
\be
\chi(x,\tau) \sim \; \frac{T^{2{x_h}} }
{(\sin(\pi T (\tau- i\; x)))^{x_h} (\sin(\pi T (\tau+ i\; x)))^{{x_h}}} \; .
\label{eq:corrfunc_qcrit_imtime}
\ee
To obtain the (retarded) susceptibility $\chi(\omega,q)$ one has to Fourier transform
this correlation function to obtain the Matsubara Green's function, 
$\chi(\omega_n,q)$ and then analytically continue it back to the real 
axis.\cite{Sachdev_book,Aronson}    The 
susceptibility has the same structure as the one obtained for the 
transverse field Ising model,\cite{Sachdev_book}     

\bea
\chi(T,\omega,q) &\sim & 
{1\over T^{26/15}} 
\label{eq:corrfunc_qcrit_realfrq}
\; \frac{\Gamma(\frac{1}{15}-i\frac{\omega+q}{4\pi T}) \; \Gamma(\frac{1}{15}-i\frac{\omega-q}{4\pi T})}{\Gamma(\frac{14}{15}-i\frac{\omega+q}{4\pi T}) \; \Gamma(\frac{14}{15}-i\frac{\omega-q}{4\pi T})}\;.
\nonumber 
\eea

This immediately implies that the dynamical susceptibility $\chi(\omega)\equiv \chi(\omega,q=0)$ shows 
$\omega/T$ scaling, 
\be
\chi(\omega,T) = \frac C {T^{26/15}} \;F(\omega/T)\;, 
\ee
where the scaling function $F(y)$ is given by 
\be
F(y) =\left(
{ \Gamma(\frac{14}{15})
\over \Gamma(\frac{1}{15})}
{ \Gamma(\frac{1}{15}-i\frac{y}{4\pi})
\over \Gamma(\frac{14}{15}-i\frac{y}{4\pi})}\right)^2
\;,
\ee
and has the following asymptotic properties:
\be
F(y) \approx 
\left\{\begin{tabular}{lr}
1 & $y \ll 1\;,$ \\ 
$\left[{\Gamma(14/15)\over \Gamma(1/15)}\right]^2
\left({y/4\pi}\right)^{-26/15} e^{i\;\pi\;13/15}\;\;\;$
& $y \gg 1\;.$  
\end{tabular}
\right.
\ee

Another quantity of interest is the 
{\em local} susceptibility, $\chi^{\rm loc}(\omega) \equiv\chi(\omega,x=0)$,  that can  also be measured  
by neutron scattering, and which can be computed to behave as
\be
\chi^{\rm local}(\omega,T) = \frac{\tilde C} {T^{11/15}}\; G(\omega/T) 
\ee
with the scaling function 
\be
G(y) = 
{ \Gamma(\frac{13}{15})
\over \Gamma(\frac{2}{15})}
{ \Gamma(\frac{2}{15}-i\frac{y}{2\pi})
\over \Gamma(\frac{13}{15}-i\frac{y}{2\pi})}
\;,
\ee
having the  asymptotic properties:
\be
G(y) \approx\left\{\begin{tabular}{lr}
1 & $y \ll 1\;,$\\ 
${\Gamma(13/15)\over \Gamma(2/15)}
\left({y/2\pi}\right)^{-11/15} e^{i\; \pi \; 11/30}\;\;\;$
 & $y \gg 1\;.$  
\end{tabular}
\right.
\ee

We have to emphasize that, although conformal invariance only holds {\em at the quantum critical point}, 
the above expressions also apply to the entire quantum critical region, $j\gg T,\omega,q \gg \Delta$, where the 
quasiparticle gap does not play an important role.

%\begin{figure}[h]
%\centering
%\includegraphics[width=6cm]{gapfugges.eps}
%\caption{\label{abra:gapfugges_g} 
%Sketch of the quasiparticle gap $\Delta$ 
%as a function of transverse field $g$.}
%\end{figure}

\section{Conclusions}

\label{sec:conclusions}

In this paper we studied dynamical properties in the gapped phases of the 
$Q$-state Potts model in the $T\to0$ limit, and in its  quantum critical behavior
for $Q=3$ (the $Q=2$ Ising case has already been studied thoroughly in the 
literature).  {\gf Deep in the gapped phases  the $T\to0$} correlation functions were found 
to show  a {\em diffusive}  character for $Q>2$. This is a consequence of the simple structure of the 
low energy \akos{quasiparticle} scattering matrices, \akos{Eq. (\ref{eq:univ_s_matrix})},  and
can be understood in a very simple way. Consider, for the sake of simplicity, 
the ferromagnetic side, $T\ll\Delta,g<g_c$. In this limit domain walls propagate as particles 
and eventually collide with other domain walls. However, due to the simple structure 
of the scattering matrix the {\em order} of the color of domains does not change under these 
collisions. 

Therefore, if we look at a given domain, then its size and position changes in time, 
but its color (orientation) does not. In other words, each domain has a typical size $\sim \xi_c$ and 
a given domain diffuses along the chain as time flows. The colors of domains far away from each other 
are uncorrelated. Therefore, the probability that the domain at time $t$ at position $x$ 
has the same color as the domain at $t=0$ and $x=0$ is approximately given by the probability 
that the domain wall that was at position $x=0$ at time $t=0$ {\em diffused} to position 
$x$ under time $t$, and is therefore proportional to 
\be
S_{\mu\mu'} \sim   {1\over\sqrt{4 \pi t D }}\; e^{-x^2/4Dt} \;,
\ee
with $D$ the {\em diffusion constant of domains}, given analytically 
by Eq.~(\ref{eq:diffus_const}).

\begin{figure}[h]
%\centering
\includegraphics[width=6cm]{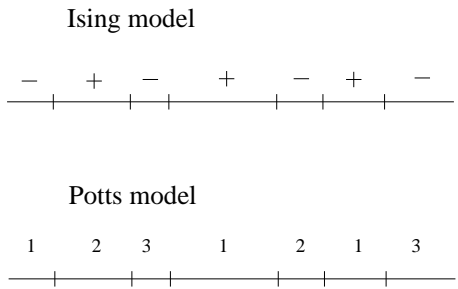}
\caption{\label{fig:domain_correlations}
Domain structure in the transverse field Ising model and in the $Q=3$ state quantum Potts model.
For $Q=2$ orientations of far away domains are correlated while for 
$Q>2$ they are not.}
\end{figure}

This simple argument, however, obviously fails for $Q=2$, 
where the colors of domains  far away from each-other  are {\em strongly correlated}, as 
illustrated  in Fig.~\ref{fig:domain_correlations}. Mathematically, 
from Eq.~(\ref{eq:Rmatrixelem_ferro}) we see that correlations between 
the colors of  two domains separated by $n$ domain walls decay exponentially 
for $Q>2$,  $C\sim e^{-n \;\ln(Q-1)}$,  while they simply oscillate for $Q=2$. 
These correlations for $Q=2$ lead to a destructive 'interference' 
and give rise to an exponential decay of the correlation functions 
for the Ising case, $Q=2$.\cite{sachdev_young} Note that the $Q=2$ Potts model is integrable\cite{integrability},
and a similar  exponential decay has also been  found recently by Altshuler\akos{, Konik} and Tsvelik in 
some other exactly integrable cases.\cite{Tsvelik}

{\gergo On the other hand, after this work was completed, we became aware of independent
work by Damle and Sachdev,\cite{Damle2} who find that diffusive
behavior appears in the correlators of the one-dimensional 
sine-Gordon model too. Rather remarkably, the decay
functions in the two cases seem to differ only in the choice of the
parameter $Q-1$, which takes a value   $Q-1 \to  -1/\cos(2\pi\eta/\gamma)$ in 
the sine-Gordon calculation,
%.  In retrospect, the sine-Gordon  model
%answer could have  also be obtained by our methods, 
although the two models  are  completely different. 
This similarity  between the two results indicates that the form of
correlation function we have found is remarkably  universal.
% and further underlines the special nature of the Ising correlations.
}

We have to emphasize that the diffusive character of the correlation  functions 
for $Q>2$ is a consequence of the simple structure (\ref{eq:univ_s_matrix}) of the scattering matrix
between quasiparticles of vanishingly small momenta.  At finite temperature, however, 
the colliding particles have finite momenta $\sim \sqrt{T\Delta/c^2}$, and therefore 
the quantum numbers of the colliding particles change with a finite probability 
in course of a collision. As a consequence, the color of the domain between the 
two colliding domain walls can {\em change} with a finite probability.
At long enough time scales, this process should lead to a decay of correlations, 
unless the correlation function  of the density of a conserved quantity is computed.\cite{Damle} 

The probability of color changing
can be estimated to be  proportional to $\sim \langle a^2 (\Delta p)^2\rangle$ 
with $\Delta p$ the typical momentum difference of the two incoming particles and $a$ the lattice constant, 
and scales  as $P_{\rm flip} \sim a^2 \Delta \; T/c^2$. A given domain 
shrinks to zero due to the motion of domain walls   approximately 
$t/\tau = \bar t$ times  during a time period  $t$, and each time it shrinks to zero 
it changes color with the above probability.

 As a result, the diffusive form of the decay function in the quantum Potts model 
 must break down at a time scale 
\be 
t_{\rm diff} \sim \tau    \frac{c^2}{ a^2 \Delta \; T}\;,
\ee
above which  correlation functions must decay exponentially.  

{\gf We also present very robust renormalization group  arguments 
to show that on the lattice the asymptotic ${\cal S}$-matrix generically takes the 
exchange form found in the perturbative regimes, and thus the diffusive 
correlation function we found should also describe the vicinity of the 
quantum critical point as $T\to 0$, $|g-g_c|\ll 1$. In this regime, however, further numerical 
calculations are needed to determine the applicability and range of validity of our 
formulas.}

Our results are not valid for 
the case $Q=4$, where 
scattering matrix cannot be computed in the way outlined in 
 Appendix \ref{sec:perturbation_theory}, because the matrices become singular 
for  $k,k'\to 0$, and cannot be inverted. As a consequence, the scattering matrix has a 
singular structure 
for $k,k'\to 0$, and a special treatment is needed. Indeed, the $Q=4$ state Potts model is 
known to have unusual thermodynamical properties due to the presence of 
a marginal operator at the critical point.\cite{marginalop} 

Also, in the present paper we neglected umklapp processes. It has been shown recently, that 
in one dimension such processes may dominate the relaxation of some currents 
that overlap with other {\em conserved} quantities.\cite{andrei_rosch} 

These processes are, however, probably not relevant 
for the correlation functions computed here, since ``spin conservation'' in course of the 
scattering process is anyway only approximate in the Potts model. These umklapp processes 
are, however,  probably important for {\em energy relaxation}, which has not been 
studied in this paper.

Finally, let us discuss  physical applications of our calculations. 
Though the 3-state quantum Potts model can 
be realized in trigonal ferromagnets in a magnetic field, and also 
recently some spin chain models have been shown to map on this model,\cite{orignac} 
it is physically not easy to find 
realizations of the 1-dimensional $Q$-state quantum Potts model. However, 
our results obtained for the paramagnetic phase are very general, and carry over to 
gapped antiferromagnets  and  spin ladders with  little modifications. 
For the spin-spin correlation function of a one $S=1$ spin Heisenberg antiferromagnet, {\em e.g.}, 
our results imply that the spin-spin correlation function decays as 
\be 
\langle {\vec S}(x_i,t)\cdot {\vec S}(0)\rangle = (-1)^i \; A\;  K(x_i,t) \; R_{Q=4}(\bar x_i ,\bar t) \;, 
\ee 
with the Feynman propagator $K$ and the decay function $R$ given by equations 
(\ref{eq:Feynmanprop_para}) and (\ref{eq:relax_fv_integralalak}). 
This result implies that the dynamical susceptibility has an approximate diffusive structure 
also at momentum $q=\pi/a$, similar to the diffusive structure that appears  
at $q=0$.\cite{Damle} However, while the $q=0$ diffusion pole follows from the 
SU(2)-invariance of the Hamiltonian and is therefore protected by symmetry, 
the $q=\pi/a$ diffusive character \akos{of the AF Heisenberg model} is only approximate, and is 
a consequence of the simple exchange form of the \akos{quasiparticle}scattering 
matrix.

{\gergo We would like to express our special  thanks to 
Kedar Damle and Subir Sachdev for showing us their preprint before  making it public, 
and F. Igl\'oi for valuable comments and discussions.} 
We would also like to thank A. Tsvelik and R. Konik for useful discussions.
This research has been supported by Hungarian Grants No. OTKA T046303, T046267, 
and NF061726.

\appendix

\section{Derivation of the scattering matrices}

\label{Appendix}

\subsection{Scattering matrices - ferromagnetic side}

To obtain the two-particle scattering matrix in leading
order  in $g$ we make the simple ansatz Eq.~(\ref{eq:2p-hullamfg_ferro})
for the two-particle wave function $|\Psi\rangle$,
\bea
|\Psi\rangle &=& 
\sum_{\theta,\theta'}
\Big\{ A_{\theta,\theta'} \sum_{i<j} \left( e^{i({k}{x_i} +
{k'}{x_j})}
|i, \theta; j,\theta' \rangle_{\tilde\mu}  \right) +
\nonumber
\\
&& B_{\theta,\theta'} \sum_{i>j} \left( e^{i({k}{x_i} +
{k'}{x_j})} | i,\theta; j, \theta' \rangle_{\tilde\mu}  \right)
\Big\} \;, \eea 
where $\tilde\mu$ is the orientation of the chain
on the far left, and $\theta$ and $\theta'$ denote the two kinks
corresponding to the two domain walls. In leading order in $g$ and
for coordinates  $|i-j|\gg 1$, $H_g$ just moves the two domain
walls independently, and $|\Psi\rangle$ is clearly  an eigenstate
of the total Hamiltonian with an eigenvalue
\begin{equation}\label{eq:append_ferro}
H | \Psi  \rangle =  \left( E_0 + \epsilon(k) +
\epsilon(k') \right) | \Psi \rangle \;,
\label{schrodinger}
\end{equation}
$E_0$ being the ground state energy, and $\epsilon(k)$ the quasiparticle energy given by
Eq.~(\ref{eq:ferro_dispersion}). However, $|\Psi\rangle$ must satisfy Eq.~(\ref{schrodinger})
also for $j=i+1$, {\em i.e.} for  nearest neighbors. Observing that the
operator $H_g =  -jg \sum_n {\tilde{P}}_n$ just flips each spin to some
other direction, we can write  in the original notation
\bea
&& \sum_n  {\tilde{P}}_n | \tilde\mu, i, \mu, i+1, \mu' \rangle =
\nonumber
\\
&&\phantom{nn}
 \;\langle \tilde\mu | {\tilde{P}} | \mu \rangle \; | \tilde\mu, i-1, \mu, i+1, \mu' \rangle +
\nonumber
\\
&& \phantom{nn}  \langle \mu | {\tilde{P}} | \mu' \rangle \;  | \tilde\mu, i, \mu, i+2, \mu' \rangle  +
\nonumber
\\
&& \phantom{nn}   \sum_{\mu'' \ne \mu} \langle \mu''|
 {\tilde{P}} | \mu \rangle \;
| \tilde\mu, i, \mu'', i+1, \mu' \rangle
+\dots \;,
\label{eq:H_1_hatasa_ferro}
\eea
where
$\theta=(\mu-\tilde\mu)\;{\rm mod}\;Q $, $\theta'=
(\mu'-\mu)\;{\rm mod}\;Q$,
and  we neglected all other terms involving more than two domain wall excitations, since these
are high up in energy.
Since the off-diagonal  matrix elements of ${\tilde{P}}_n$ are all equal
to $1/Q$, we can write this in the 'kink'  representation as:
\bea
&&\sum_n {\tilde{P}}_n |i,\theta ;\; i+1, \theta' \rangle_{\tilde\mu} =
\label{eq:H_1_hatasa_2_ferro}
\\
&& \frac1Q \;\Bigl[| i-1,\theta ;\; i+1, \theta' \rangle_{\tilde \mu} +
 | i,\theta ;\; i+2, \theta' \rangle_{\tilde\mu} +
\nonumber
\\
&&
 \sum_{\tilde\theta,\tilde\theta'}
\hat \delta_{\theta+\theta'}^{\tilde\theta+\tilde\theta'}
\; (1-\delta_{\theta}^{\tilde\theta'})\; | i,\tilde\theta' ;\;
i+1, \tilde\theta \rangle_{\tilde\mu}
\Bigr]
+\dots\;,
\nonumber
\eea
where
$\hat \delta$ denotes the
Kronecker delta modulo $Q$. Projecting  out the
 $i+1=j$ component of the Schr\"odinger equation Eq.~(\ref{schrodinger})
we obtain the following constraint for the coefficients
$A_{\theta,\theta'}$ and $B_{\theta,\theta'}$,
\bea
&&\sum_{\tilde\theta,\tilde\theta'} \left\lbrack
\delta_{\theta}^{\tilde\theta}
\delta_{\theta'}^{\tilde\theta'} + {1\over e^{i k'}+ e^{-ik} }
                   \hat \delta_{\theta+\theta'}^{\tilde\theta+\tilde\theta' }
                                      (1- \delta_{\theta}^{\tilde\theta'}) \right\rbrack B_{\tilde\theta',\tilde\theta} =
\nonumber
\\
&&-\sum_{\tilde\theta,\tilde\theta'} \left\lbrack
\delta_{\theta}^{\tilde\theta}
\delta_{\theta'}^{\tilde\theta'}
+ {1 \over e^{ik}+ e^{-ik'}}
  \hat \delta_{\theta+\theta'}^{\tilde\theta+\tilde\theta'}
(1- \delta_{\theta}^{\tilde\theta'}) \right\rbrack A_{\tilde\theta,\tilde\theta'}
\nonumber
\;.
\label{eq:Smatrix_osszefugges}
\eea

This equation can clearly be inverted to give the two-particle
$\cal S$-matrix in leading order in $g$, but the solution
is rather complicated even for $Q=3$.
Eq.~(\ref{eq:Smatrix_osszefugges}) simplifies, however, in the limit
$k,k'\to 0 $, relevant for very small temperatures,
$T \ll \Delta$.

For $Q=3$ we obtain in this way
\begin{equation}\label{eq:smatrix_limeszben_ferro}
{\cal S}_{\theta,\theta'}^{\tilde\theta,\tilde\theta'}(k,k'\to0)= (-1) \;
\delta_{\theta}^{\tilde\theta'} \;\delta_{\theta'}^{\tilde\theta}
\;.
\end{equation}

We remark here that the above result holds for any $Q\ne4$. The
$Q=4$ case, however, seems to be special: then the operator in
front of the coefficients $ A_{\tilde\theta,\tilde\theta'}$ and
$B_{\theta,\theta'}$  in Eq.~(\ref{eq:Smatrix_osszefugges}) has
zero eigenvalues for $k=k'=0$, the inversion is problematic, and
the $\cal S$-matrix does not take the form (\ref{eq:smatrix_limeszben_ferro}). 

\subsection{Scattering matrices - paramagnetic side}

To obtain the two-particle scattering matrix in leading order in $1/g$ we follow 
similar steps as for the ferromagnetic case. The ansatz for the two-particle wave function
can be written as follows:
\bea
|\tilde{\Psi}\rangle &=& \sum_{\lambda,\lambda'}
\Big\{ A_{\lambda,\lambda'} \sum_{i<j} \left( e^{i({k}{x_i} + {k'}{x_j})}
|i,\lambda; j,\lambda' \rangle  \right) + \nonumber
\\
&& B_{\lambda,\lambda'} \sum_{i>j} \left( e^{i({k}{x_i} +
{k'}{x_j})} | i,\lambda; j, \lambda' \rangle  \right) \Big\} \;.
\eea

For $|i-j| \gg 1$ $ |\tilde{\Psi}\rangle $ must satisfy a similar two-particle
Schr\"odinger equation as (\ref{eq:append_ferro}), and so it must be valid for $j=i+1$ too.
Let us now calculate the effect of 
$H_{\rm ferro}=- j\sum_{n} \sum_{\mu}{\tilde{P}}^{\mu}_n  {\tilde{P}}^{\mu}_{n+1}\;$ on the $i+1=j$
term of $ |\tilde{\Psi}\rangle $. If we neglect the high energy terms with multiple particles we get
\bea
\sum_{n} \sum_{\mu}{\tilde{P}}^{\mu}_n  {\tilde{P}}^{\mu}_{n+1} \; |i,\lambda;i+1,\lambda'\rangle = \\
1/Q \; |i-1,\lambda;i+1,\lambda'\rangle + \nonumber \\ 
1/Q \; |i,\lambda;i+2,\lambda'\rangle + \nonumber \\ 
\sum_{\tilde{\lambda} \tilde{\lambda}'} \sum_{\mu} 
\langle \tilde{\lambda}' | {\tilde{P}}^{\mu}| \lambda \rangle  \langle \tilde{\lambda}|{\tilde{P}}^{\mu}| \lambda'\rangle |i,\tilde\lambda';i+1,\tilde\lambda\rangle \nonumber \;.
\eea

Substituting this to the two particle Schr\"odinger equation leads to a similar constraint for the coefficients as in the ferromagnetic case:
\bea
&&\sum_{\tilde\lambda,\tilde\lambda'} \left\lbrack
\delta_{\lambda}^{\tilde\lambda}
\delta_{\lambda'}^{\tilde\lambda'} + {1\over e^{i k'}+ e^{-ik} }
                   M_{\lambda \lambda'}^{\tilde\lambda \tilde\lambda' }
                                       \right\rbrack B_{\tilde\lambda',\tilde\lambda} =
\\
&&-\sum_{\tilde\lambda,\tilde\lambda'} \left\lbrack
\delta_{\lambda}^{\tilde\lambda}
\delta_{\lambda'}^{\tilde\lambda'}
+ {1 \over e^{ik}+ e^{-ik'}}
  M_{\lambda \lambda'}^{\tilde\lambda \tilde\lambda'}
 \right\rbrack A_{\tilde\lambda,\tilde\lambda'}
\nonumber
\;,
\label{eq:Smatrix_osszefugges_para}
\eea
where 
\bea
M_{\lambda \lambda'}^{\tilde\lambda \tilde\lambda'}=\sum_{\mu} 
\langle \tilde{\lambda}' | {\tilde{P}}^{\mu}| \lambda \rangle  
\langle \tilde{\lambda}|{\tilde{P}}^{\mu}| \lambda'\rangle .
\label{eq:M_matrixdef}
\eea

Similar to the ferromagnetic case, in the limit of vanishing quasiparticle momenta
$k,k' \to 0$, this equation can be solved to obtain the $\cal S$-matrix for $Q=3$:
\begin{equation}\label{eq:smatrix_limeszben_para}
{\cal S}_{\lambda,\lambda'}^{\tilde\lambda,\tilde\lambda'}(k,k'\to0)= (-1) \;
\delta_{\lambda}^{\tilde\lambda'} \;\delta_{\lambda'}^{\tilde\lambda}
\;.
\end{equation}

{\bff 
\section{Computation of the relaxation functions assuming a diagonal $\cal S$-matrix for $Q=3$}
\label{app:integr}

\subsection{\akos{Diagonal} case: Ferromagnetic side}

Let us compute the correlation function $C_{\mu\mu'}^{\rm ferro} (x,t)$
 defined as the  probability that the domain containing  $(0,0)$ has the orientation 
$\mu$ while that at $(x,t)$  has orientation $\mu'$. Note that $C_{\mu\mu'}$ is 
defined in terms of the  $P^\mu$'s rather than the $\tilde P^{\mu}$'s. 

We can calculate this probability by just keeping track of the  number $N_+$ and $N_-$  of 
kink excitations (domain walls) with quantum numbers $+$ and $-$, respectively,
that cross the line between the points $(0,0)$ and $(x,t)$. 
The points $(0,0)$ and $(x,t)$ have  the same domain orientation if $N_+$ 
equals $N_-$ up to modulo three.

Thus the correlation function is
\bea 
&&C_{\mu\mu'}^{\rm ferro} (x,t)  = \frac 13 \sum_n  P_n[(0,0) \to(x,t)] \times
\nonumber 
\\
&& 
 \phantom{nnn} \times \sum_k {n \choose k} \frac{1}{2^n} \hat\delta(2k-n + \Delta \mu) \;, 
\eea
where $ P_n[(0,0) \to(x,t)] $ is the probability that the line $(0,0) \to(x,t)$ cuts precisely 
$n$ domain walls, $k$ is the number of upsteps, $\hat\delta(m)$ is the discrete delta function 
modulo 3, and $\Delta \mu = \mu' -\mu$.
The probability that we cut the path of $n$ particles when moving from $(0,0)$ to $(x,t)$ can be computed 
in terms of cutting none:
\be
 P_n((0,0) \to(x,t) ) = {M \choose n} \bar P ^{M-n} (1-\bar P)^n \;,
\ee
where $M$ is the total number of particles and
\bea
\bar P& = &P({\rm will\;not\;cut}\; \nu' th )  =  \int_{-L/2}^{0} \frac{dx_\nu}{L} \int_{-\infty}^{\frac{x-x_\nu}{t}} dv_\nu P(v_\nu)
\nonumber \\
&& \phantom{nnn} +  \int^{L/2}_{0} \frac{dx_\nu}{L} \int^{\infty}_{\frac{x-x_\nu}{t}} dv_\nu P(v_\nu) \;.
\eea
which can be calculated to get
\be
P({\rm will\;not\;cut}\; \nu' th ) = 1 - \frac{\bar t }{M} \left[ \frac{1}{\sqrt{\pi}} e^{-u^2} + u \; {\rm erf} (u) \right]\;,
\label{eq:cutting_none}
\ee 
with $u$ the dimensionless velocity introduced earlier, $u = \bar x/\bar t$. 

Using then the representation of the  delta function 
\be 
\hat\delta(m) = {\rm Re} \left[\frac{2}{3} e^{i\frac{2\pi}{3} m}\right]  + \frac{1}{3} \;,
\ee
all summations above can be evaluated to yield 
\be
C_{\mu\mu'}^{\rm ferro} (x,t) =  \frac{1}{9} 
+ \frac{1}{3} \left( \delta_{\mu \mu'}-\frac{1}{3}\right) e^{- \frac{3}{2}\bar t G(u)}, 
\ee
leading to the relaxation function given in the main text, Eq.~(\ref{eq:R_int^ferro}). 

\subsection{\akos{Diagonal} case: Paramagnetic side}

Two differences arise when we use the diagonal $\cal S$-matrix, 
Eq.~(\ref{eq:integr_s_matrix}). The first one is that the 
sequence of the colors now does change in course of the scattering process, while the quantum number of a given 
line does not. The second change is that the number of the (-1) factors picked up by the wave function 
now depends on the quantum numbers of two colliding particles. This second change implies 
that, after averaging over the color of the other quasiparticles, one gets identically $0$ if 
the particle created by the operator ${\tilde P}^\mu$ collides with any other particle. 
Thus the relaxation function is simply the probability that the created particle 
propagating from point $(0,0)$ to $(x,t)$ collides with no other particle.

The probability that the $\nu$'th particle will not collide with the injected one is given
by Eq.~(\ref{eq:cutting_none}).

Thus the probability that the injected particle does not collide with any of the 
already existing quasiparticles simply reads
\bea
&&R_{\rm integrable}^{\rm para}(\bar x,\bar t) = \prod_\nu P({\rm will\;not\;cut}\; \nu' th )=
\nonumber 
\\
&&\phantom{nnn} =  
\left(1 -\frac{\bar t G(u)}{M} \right)^M \to e^{-\bar t G(u)}\;,
\eea

resulting in the relaxation function given by Eq~(\ref{eq:R_int^para}).

}

\section{Relation between the effetive Hamiltonian and the scattering matrix}
\label{appendix:RG} 

Already simple renormalization group arguments suggest
that the  ${\cal S}$-matrices  above describe 
 the scattering of quasiparticles with vanishing momenta 
for {\em any} coupling $g\ne g_c$ {\gf and $T \ll \Delta$}: 
We know very well that the coupling $g$ is relevant for 
$g> g_c$ and scales to $g\to \infty$  under the RG flow , 
while for $g<g_c$  it is irrelevant and scales to $g\to 0$. 
The asymptotic (long wavelength)  dynamical properties, however, 
remain  invariant under the renormalization group, and therefore 
the scattering matrix of quasiparticles with vanishing momenta 
obtained  by simply performing perturbation theory in $1/g$ (or $g$) 
around the trivial fixed point Hamiltonians 
must coincide with the {\em exact} scattering 
in both phases.

%We firmly believe that the scattering matrix above is {\em the 
%universal S-matrix} of the quantum Potts model {\em on a lattice (with a 
%finite momentum cut-off).} 

One can refine the renormalization group argument above to 
show that,  apart from some very  special points, the $\cal S$-matrix of the 
quantum Potts model on a lattice should always  take on the simple form, 
Eq.~(\ref{eq:univ_s_matrix}) for vanishing quasiparticle momenta as follows.  
Let us suppose that we have 
a high energy cut-off $\Lambda$,  larger than the gap, $\Lambda > \Delta$. 
Performing  a renormalization group transformation down to a length scale 
$b \gg \Delta^{-1}$, we obtain a local Hamiltonian 
for the elementary  excitations ('dressed' local flips and kinks for $g>g_c$ and 
$g<g_c$, respectively),  that we can simply 
construct based upon symmetry considerations and power counting:
\bea
 H_{\rm eff} & = &   -\sum_i \frac{c^2}{2\Delta}\; \frac{\partial}{{\partial x_i}^2}  + \sum_{i<j}
u^{\lambda_i', \lambda_j'}_{\lambda_i, \lambda_j} \;\delta(x_i-x_j) 
\nonumber
\\
& + & \mbox{irrelevant terms}\;,  
\eea
where $x_i$ denotes the coordinate of the $i$'th quasiparticle.
Here the second interaction  term is always relevant, and it is this term that determines 
the structure of the $\cal S$-matrix for vanishing momenta.
 For the sake of simplicity, let us restrict ourselves to the 
most interesting $Q=3$ case. Then $\lambda$ is just a chirality label, 
$\lambda = \pm$, and the interaction 
matrix  $u^{\lambda_i', \lambda_j'}_{\lambda_i, \lambda_j}$ can be 
characterized through three dimensionless parameters,
$u_1 \equiv u^{++}_{++} =  u^{--}_{--}$,
 $u_2 \equiv u^{+-}_{+-} =  u^{-+}_{-+}$, and 
 $u_3 \equiv u^{+-}_{-+} =  u^{-+}_{+-}$. These parameters are 
 dimensionless functions of $\Delta \;b$ and $\Delta/\Lambda$
and can be written as 
\be
u_\alpha = u_\alpha\left(\Delta b, \frac \Delta \Lambda\right)\;,
\;\;\;\;\;\;\;\;\alpha=1,2,3\;.
\ee
 In the $b\to \infty $ (large wavelength) limit these must 
scale  to  constants that only depend on the ratio
$\Delta/\Lambda$,
\be
\lim_{b\to \infty} u_\alpha\left(\Delta b, \frac \Delta \Lambda \right)
 = u_\alpha\left(\frac \Lambda \Delta\right)\;.
\ee
These three numbers determine then the  asymptotic form of the $\cal S$-matrix.
It is then a simple matter to show that the 
$\cal S$-matrix generated by this effective interaction {\em always} assumes the 
form Eq.~\ref{eq:univ_s_matrix},\cite{Rapp_future} excepting the special case when
\be
u_2^2 = u_3^2\;.
\label{eq:integrabilitycondition}
\ee
In this latter case the interaction is singular 
(vanishes in one of the scattering channels) and one obtains 
an $\cal S$-matrix that is  diagonal
for vanishing quasiparticle momenta, as given by Eq.~(\ref{eq:integr_s_matrix}).   
This $\cal S$-matrix coincides with the one that one obtains by 
requiring {\em integrability}, i.e. by requring that the scattering matrices satisfy  
the Yang-Baxter relations.\cite{Read} 
Thus  Eq.~(\ref{eq:integrabilitycondition})
can be viewed as an {\em integrability condiction}. 
 However, the $Q=3$ state  Potts model 
is {\em not} integrable away from the critical point.

Although there is no obvious reason for that, 
nevertheless,  it might be  possible that 
Eq.~(\ref{eq:integrabilitycondition}) is satisfied if one  
removes the cut-off from the theory, {\em i.e.} if one takes the limite
$\Lambda/\Delta \to \infty$. 
Indeed, the scattering matrix (\ref{eq:integr_s_matrix}) 
is only obtained  in a continuum field theory approach, when 
one {\em removes} the cut-off and also {\em assumes} integrability.\cite{Read,Zamolodchikov,Swieca}.

However, in a theory {\em with a cut-off} there 
is not a single reason why the integrability 
condition, Eq.~(\ref{eq:integrabilitycondition}), should be 
satisfied. In fact, our perturbative expansion just proves
that Eq.~(\ref{eq:integrabilitycondition})  is not satisfied 
in the large $g$ and small $g$ limits.

These arguments very convincingly support that
the $\cal S$-matrix of the quantum Potts model on a lattice \akos{in its gapped phases 
at $T \to 0$} always takes on the the form 
(\ref{eq:univ_s_matrix}) in the limit of vanishing momenta, 
and the low energy fixed point theory is simply generically 
 {\em not integrable}.\cite{Rapp_future}
It is an interesting question if there exists still a cross-over regime for 
$\Delta \ll \Lambda$, {\em i.e.}, if there is some regime where  
 for intermediate energies  $\Delta \ll c^2 q^2/\Delta \sim T  > T^*$,  
the diagonal $\cal S$-matrix is adequate, but the discussion of this 
requires extensive numerical studies and is certainly beyond the scope of the present 
paper.\cite{Rapp_future}


\begin{thebibliography}{99}
\bibitem{integrability} The $Q>2$ Potts model is believed to be only integrable at the critical point.
\end{thebibliography}
\end{document}